\documentclass[12pt]{article}
\usepackage[utf8]{inputenc}
\pdfoutput=1

\usepackage[paper=letterpaper,margin=1in]{geometry}

\usepackage{amsmath,amssymb,amsfonts,epsfig,cite,setspace,bigstrut,longtable,array,url,xcolor,microtype,mathtools,color}
\definecolor{darkblue}{rgb}{0.0,0.0,0.3} 		
\mathtoolsset{centercolon}
\usepackage{caption}
\usepackage{subcaption}
\usepackage[unicode=true,
bookmarks=true,bookmarksnumbered=false,bookmarksopen=false,
breaklinks=false,pdfborder={0 0 0},pdfborderstyle={},backref=false,colorlinks=true]
{hyperref}
\hypersetup{pdftitle={Machine-Learning for Calabi--Yau Metrics},
	pdfauthor={},
	linkcolor=darkblue, urlcolor=darkblue, citecolor=darkblue, filecolor=black, linktoc=page}

\allowdisplaybreaks

\usepackage{lipsum}

\let\OLDthebibliography\thebibliography
\renewcommand\thebibliography[1]{
  \OLDthebibliography{#1}
  \setlength{\parskip}{0pt}
  \setlength{\itemsep}{0pt plus 0ex}
}

\makeatletter
\g@addto@macro\bfseries{\boldmath}
\makeatother

\let\originalleft\left
\let\originalright\right
\renewcommand{\left}{\mathopen{}\mathclose\bgroup\originalleft}
\renewcommand{\right}{\aftergroup\egroup\originalright}

\definecolor{purple}{cmyk}{0,0.8,0,0.4}
\definecolor{db}{cmyk}{1,0.1,0.2,0.6}
\definecolor{dg}{cmyk}{1,0,1,0.7}
\definecolor{bl}{cmyk}{1,0.7,0.5,0.2}
\definecolor{yl}{rgb}{0.2,0.7,0.2}
\definecolor{bl2}{cmyk}{0.7,0.4,0,0.5}
\definecolor{red2}{cmyk}{0,1,1,0.8}
\definecolor{red3}{cmyk}{0,0.7,1,0.7}
\definecolor{gr2}{cmyk}{1,0.2,0.7,0.6}
\definecolor{nb}{rgb}{0.1,0.1,0.5}
\definecolor{ng}{rgb}{0,0.8,0}
\definecolor{brown}{rgb}{0.6,0.3,0.2}
\definecolor{newred}{cmyk}{0,1,1,1}

\DeclareMathOperator{\tr}{tr}					
\newcommand{\dd}{\mathrm{d}} 								
\newcommand{\del}{\partial} 								
\newcommand{\delb}{\bar{\partial}} 								
\newcommand{\ee}{\mathrm{e}} 								
\newcommand{\ii}{\mathrm{i}} 								
\newcommand{\Ob}{\overline{\Omega}} 								
\newcommand{\detg}[1]{g^{(#1)}}
\newcommand{\detgp}[1]{\hat{g}^{(#1)}}
\newcommand{\gk}{g^{(k)}_{a\bar b}}

\newcommand{\hmat}{h^{\alpha\bar\beta}}

\begin{document}
	
\newcommand{\nn}{\nonumber}
\newcommand{\ch}{\rm Ch}
\newcommand{\comment}[1]{}
\newcommand{\eqspace}{\mathrel{\phantom{=}}{}} 				

\newcommand{\cM}{{\cal M}}
\newcommand{\cW}{{\cal W}}
\newcommand{\cN}{{\cal N}}
\newcommand{\cH}{{\cal H}}
\newcommand{\cK}{{\cal K}}
\newcommand{\cZ}{{\cal Z}}
\newcommand{\cO}{{\cal O}}
\newcommand{\cB}{{\cal B}}
\newcommand{\cC}{{\cal C}}
\newcommand{\cD}{{\cal D}}
\newcommand{\cE}{{\cal E}}
\newcommand{\cF}{{\cal F}}
\newcommand{\cR}{{\cal R}}
\newcommand{\cT}{{\cal T}}
\newcommand{\cV}{{\cal V}}
\newcommand{\IA}{\mathbb{A}}
\newcommand{\IB}{\mathbb{B}}
\newcommand{\IP}{\mathbb{P}}
\newcommand{\IQ}{\mathbb{Q}}
\newcommand{\IH}{\mathbb{H}}
\newcommand{\IR}{\mathbb{R}}
\newcommand{\IC}{\mathbb{C}}
\newcommand{\IF}{\mathbb{F}}
\newcommand{\IM}{\mathbb{M}}
\newcommand{\II}{\mathbb{I}}
\newcommand{\IZ}{\mathbb{Z}}
\newcommand{\sym}{{\rm Sym}}

\newcommand{\tmat}[1]{{\tiny \left(\begin{matrix} #1 \end{matrix}\right)}}
\newcommand{\mat}[1]{\left(\begin{matrix} #1 \end{matrix}\right)}
\newcommand{\diff}[2]{\frac{\partial #1}{\partial #2}}
\newcommand{\gen}[1]{\langle #1 \rangle}
\newcommand{\ket}[1]{| #1 \rangle}
\newcommand{\jacobi}[2]{\left(\frac{#1}{#2}\right)}

\def\acts{\curvearrowright}

\newcommand{\drawsquare}[2]{\hbox{%
\rule{#2pt}{#1pt}\hskip-#2pt
\rule{#1pt}{#2pt}\hskip-#1pt
\rule[#1pt]{#1pt}{#2pt}}\rule[#1pt]{#2pt}{#2pt}\hskip-#2pt
\rule{#2pt}{#1pt}}
\newcommand{\fund}{\raisebox{-.5pt}{\drawsquare{6.5}{0.4}}}
\newcommand{\antifund}{\overline{\fund}}

\newtheorem{theorem}{\bf THEOREM}
\def\thetheorem{\thesection.\arabic{theorem}}
\newtheorem{proposition}{\bf PROPOSITION}
\def\thetheorem{\thesection.\arabic{proposition}}
\newtheorem{observation}{\bf OBSERVATION}
\def\thetheorem{\thesection.\arabic{observation}}

\def\theequation{\thesection.\arabic{equation}}
\newcommand{\setall}{\setcounter{equation}{0}
        \setcounter{theorem}{0}}
\newcommand{\setequation}{\setcounter{equation}{0}}

~\\
\vskip 1cm

\begin{center}
{\Large \bf Machine Learning Calabi--Yau Metrics}
\end{center}
\medskip

\vspace{.4cm}
\centerline{
{\large Anthony Ashmore},$^{1,2}$
{\large Yang-Hui He},$^{2,3,4}$
{\large Burt A.~Ovrut}$^1$
}
\vspace*{3.0ex}

\begin{center}
{\it
{\small

${}^{1}$
Department of Physics, University of Pennsylvania,
Philadelphia, PA 19104, USA \\

${}^{2}$
Merton College, University of Oxford, OX1 4JD, UK \\

${}^{3}$
Department of Mathematics, City, University of London, EC1V 0HB, UK\\

${}^{4}$
School of Physics, NanKai University, Tianjin, 300071, P.R.~China \\

\vspace*{1.5ex}
\qquad 
{\rm 
\url{aashmore@sas.upenn.edu},
\url{hey@maths.ox.ac.uk},
\url{ovrut@upenn.edu}}

}}
\end{center}

\begin{abstract}
We apply machine learning to the problem of finding numerical Calabi--Yau metrics. Building on Donaldson's algorithm for calculating balanced metrics on K\"ahler manifolds, we combine conventional curve fitting and machine-learning techniques to numerically approximate Ricci-flat metrics. We show that machine learning is able to predict the Calabi--Yau metric and quantities associated with it, such as its determinant, having seen only a small sample of training data. 
Using this in conjunction with a straightforward curve fitting routine, we demonstrate that it is possible to find highly accurate numerical metrics much more quickly than by using Donaldson's algorithm alone, with our new machine-learning algorithm decreasing the time required by between one and two orders of magnitude.
\end{abstract}

\def\ib{\bar{\imath}}
\def\jb{\bar{\jmath}}
\def\zb{\bar{z}}
\def\xb{\bar{x}}
\def\ab{\bar{\alpha}}
\def\bb{\bar{\beta}}
\def\gb{\bar{\gamma}}
\def\db{\bar{\delta}}
\def\sb{\bar{s}}
\newcommand{\vcy}[1]{\text{Vol}_{\text{CY}}}
\newcommand{\vk}[1]{\text{Vol}_{\text{K}}}

\newpage

\tableofcontents

\newpage

\section{Introduction}

The promise of string theory as a unified theory of everything rests on the belief that it can reproduce the known physics in our universe. In particular, at low energies it must reduce to the Standard Model. The first, and perhaps still the most promising, way to produce string models with realistic low-energy physics is to compactify the $\text{E}_8\times\text{E}_8$ heterotic string on a Calabi--Yau threefold~\cite{Candelas:1985en}. As it stands today, there are a number of viable heterotic models that lead to three generations of quarks/leptons with realistic gauge groups and the correct Higgs structure~\cite{Braun:2005ux,Bouchard:2005ag, Braun:2005nv,Braun:2005bw,Anderson:2009mh,Anderson:2011ns,Anderson:2012yf,Anderson:2013xka,Ovrut:2015uea,Faraggi:1989ka,Cleaver:1998saa}, with more predicted to exist~\cite{Constantin:2018xkj}.

Despite this progress, one should not lose sight of the necessary requirement that such vacua must satisfy; namely, that their observable properties be consistent with all known low energy phenomenology and properties of particle physics. To do this, one must explicitly perform top-down strings computations of observable quantities and compare the results with the experimental data. Within the context of~\cite{Braun:2005nv}, for example, the masses of  the gauge bosons and the Higgs mass have been computed to one-loop accuracy using an explicit renormalization group calculation from the compactification scale, with the results shown to be accurate~\cite{Ambroso:2009jd,Ambroso:2010pe,Ovrut:2012wg,Ovrut:2014rba,Ovrut:2015uea}. It was also demonstrated in this model that all supersymmetric sparticle masses are above their present experimental lower bounds. However, the values of the various dimensionful and dimensionless couplings of the low-energy theory -- for example, the Yukawa couplings, the gauge coupling parameters and so on -- have not been explicitly calculated to date.
Among the many such quantities one would like to compute from a top-down string model, of particular interest are the Yukawa couplings. With these in hand, one could make a concrete prediction for the masses of elementary particles from string theory.

Generic discussions of the mathematical structure of Yukawa couplings within the context of heterotic compactifications have been presented in~\cite{Braun:2006me}. Unfortunately, it is not currently possible to compute these couplings explicitly in general. To do so requires finding the gauge-enhanced Laplacian on a Calabi--Yau threefold with a holomorphic vector bundle, using this to compute the harmonic representatives of various sheaf cohomologies and then integrating a cubic product of these harmonic forms over the manifold. Unfortunately, there is no known analytic expression for the metric on a Calabi--Yau manifold, nor does one know the analytic form of the gauge connection on the vector bundle. Hence, it is presently impossible to determine the required harmonic one-forms analytically.  

A number of previous works have tried to tackle this problem numerically. Building on the seminal work of Donaldson~\cite{donaldson1,donaldson2}, there are now algorithms that approximate Ricci-flat metrics on K\"ahler manifolds and solve the hermitian Yang--Mills equations~\cite{Douglas:2006rr,Douglas:2006hz,Headrick:2005ch,Headrick:2009jz,Braun:2007sn,Braun:2008jp,Anderson:2010ke,Anderson:2011ed}. In principle, once one has the Ricci-flat metric and the gauge connection, one can find the normalized zero modes of various Laplacians on the compactification manifold and then, as stated above,  compute the Yukawa couplings.
Despite focussed work on this topic, this goal has not yet been achieved. The current state-of-the-art allows numerical calculations of the metric, the gauge connection and the eigenmodes of the ``scalar'' Laplacian (the Laplacian without the gauge connection acting on functions). There is no conceptual barrier to extending this numerical approach to the full problem of computing zero modes of gauge-coupled Laplacians. However, there is a very serious technical barrier. 
Moving away from simple Calabi--Yau manifolds, such as the quintic threefold, to non-simply connected Calabi--Yau manifolds with discrete symmetries and complicated gauge bundles -- such as those referenced above -- greatly increases both the time and computational power needed. A natural question is whether new computational techniques, such as ``machine learning'', might be useful in reducing the time and resources required for these more phenomenologically realistic vacua.

Recently, there has been a great amount of interest in applying techniques of machine learning to string theory, as pioneered in \cite{He:2017aed,He:2017set,Krefl:2017yox,Ruehle:2017mzq,Carifio:2017bov}.
In particular, methods of machine learning have been applied to various ``Big Data'' problems in string compactifications, such as string vacua, the AdS/CFT correspondence, bundle cohomology and stability, cosmology and beyond \cite{Carifio:2017nyb,Liu:2017dzi,Cohen:2017exh,Demir:2018iqo,Wang:2018rkk,Hashimoto:2018ftp,Bull:2018uow,Bull:2019cij,Cole:2018emh,Mutter:2018sra,Jinno:2018dek,Rudelius:2018yqi,Halverson:2018cio,Cunningham:2018sdj,Klaewer:2018sfl,Demirtas:2018akl,Brodie:2019dfx,Comsa:2019rcz}, as well as the structure of mathematics
\cite{He:2018jtw,Jejjala:2019kio,he2019learning,He:2019vsj}. The idea of this present work is to apply these same methods to see whether they are able to increase the accuracy and/or reduce the time and cost of numerical calculations, specifically of the Calabi--Yau metric on generic threefolds. As we will see, machine learning does appear to have a part to play in this story. 

First, we show that machine learning algorithms can ``learn'' the data of a Calabi--Yau metric. More specifically, Donaldson's algorithm involves a choice of line bundle, whose sections provide an embedding of the Calabi--Yau within projective space. The choice of line bundle fixes the degree $k$ of the polynomials that appear in an ansatz for the K\"ahler potential. As $k$ increases, the numerical metric becomes closer to Ricci-flat and the algorithm increases in both its run-time and resource requirements. For clarity, we will introduce our machine learning algorithm within the context of 
the determinant of the Calabi--Yau metric -- a single function rather than the nine components required to express the complete metric. The determinant is also of interest in its own right, since it is necessary to compute the so-called $\sigma$-measure which determines how close the metric is to being Ricci-flat. We will show that given the data of the determinant of the metric for low values of $k$, our machine-learning model can predict the determinant corresponding to higher values of $k$ (that is, closer to the actual Ricci-flat metric). However, this calculation needs to be ``seeded'' with {\it some} values of the determinant at larger values of $k$; in other words, this is a supervised learning problem. 

Unfortunately, having to ``seed'' the calculation with values of the determinant at larger $k$ -- which must be computed using Donaldson's algorithm -- greatly increases the run-time required. Ideally, one would like to take a preliminary numerical approximation to the determinant and improve on its accuracy without needing to input any data for larger values of $k$. There are a number ways one might go about this. In this paper, we use a simple extrapolation based on curve fitting to predict how the determinant behaves at larger values of $k$, leaving more complicated methods to future work. We show that this curve fitting algorithm significantly reduces the time required to compute the determinant to higher accuracy. Be that as it may, although faster than using the above machine learning calculation, curve extrapolation is still rather time and resource expensive.

To overcome this problem, we combine both of these approaches: we use the extrapolated data from curve fitting to seed a supervised learning model. Remarkably, we find that this combination of the two algorithms is able to predict the values of the determinant much more quickly than either of the approaches individually. We compare the accuracy and run-time of this combined model with extrapolation and supervised learning individually, as well as Donaldson's algorithm. We will see that one does not sacrifice much in the way of accuracy, but gains tremendously in speed. In particular, we will demonstrate a factor of roughly 75 speed-up over Donaldson's algorithm alone. 

As stated above, for clarity we present this combined algorithm within the context of calculating the determinant of the metric. We emphasise, however, that these results are immediately applicable to numerically computing the full Calabi--Yau metric. We will show this explicitly in the penultimate section of this paper.
This combined algorithm -- using Donaldson's method to compute the Calabi--Yau metric for low values of $k$, combined with curve fitting to compute a small sample of training data and finally machine learning to predict the metric for the remaining points -- is the main result of this paper. It provides a factor of 50 speed-up over using Donaldson's algorithm alone.

We plan to show:

\noindent I) Donaldson's algorithm can be pushed to greater accuracy using Mathematica's fast linear algebra routines. However, this remains time and resource intensive, both scaling factorially as we increase $k$. Our aim is to use machine learning to mitigate these problems.\\	

\noindent II) Focussing on the determinant of the Calabi--Yau metric for clarity:
\begin{itemize}
	\item Using supervised learning, a machine-learning algorithm (ML) can be trained to predict properties of a Calabi--Yau metric, specifically the determinant. This will show that the geometry of Calabi--Yau manifolds is amenable to the techniques of machine learning, at least in principle.
	\item Unfortunately, the nature of supervised learning means that we need some sample data for whatever we are trying to predict. To side-step this, we use a straightforward curve-fitting analysis to extrapolate from lower accuracy, easily computable data to higher accuracy data that is otherwise very time consuming to obtain via Donaldson's algorithm.
	\item Curve fitting for a larger number of data sets is also time consuming. To avoid the shortcomings of both the machine-learning and curve-fitting approaches, we combine them. Curve fitting provides an easy way to obtain accurate values of the determinant that can then be used to train machine-learning via supervised learning. The curve fitting needs to be done on only a small sample of the total data since the ML needs only a small training set. Together, this allows one to compute the metric data many times more quickly than Donaldson's algorithm alone, with a factor of 75 speed-up for the determinant. 
\end{itemize}

\noindent III) For the complete Calabi--Yau metric:

\begin{itemize}
\item The combined algorithm presented for predicting the determinant of the metric, that is, using both machine learning and curve fitting, will be shown to be applicable for computing the {\it complete} Calabi--Yau metric. We show that this allows one to compute the complete metric data many times more quickly than using Donaldson's algorithm alone, with a speed-up by a factor of 50 or so for the full metric. 
\end{itemize}

We begin in Section 2 with an overview of Donaldson's algorithm for approximating Calabi--Yau metrics, with a more detailed discussion presented in Appendix \ref{app:donaldson}. In Section 3 we outline the general ideas of machine learning and the specific kind of machine learning we will be using, namely supervised learning. We then discuss how supervised learning can be applied to predict the data of the approximate Calabi--Yau metric. In Section 4, we outline how to extrapolate higher-accuracy data from lower-accuracy data via curve fitting, and we combine this with supervised learning in Section 5. Section 6 is devoted to showing that this combined algorithm, that is, using machine learning along with curve fitting a small number of training points, is directly applicable to the complete nine-component Calabi--Yau metric. We finish the text with a discussion of future work.
The appendices contain a detailed discussion of Donaldson's algorithm, a description of our numerical routine implemented in Mathematica and a rewriting of various error measures, a discussion of the machine-learning algorithm we use and finally, as a sanity check, we show that machine learning cannot be replaced by simple regression.

\section{Calabi--Yau metrics and Donaldson's algorithm}

We begin with a review of Calabi--Yau metrics, Yukawa couplings and Donaldson's algorithm for finding numerical metrics on Calabi--Yau manifolds~\cite{donaldson1}. A more detailed discussion for the particular case of the Fermat quintic is included in Appendix \ref{app:donaldson}.

Let $X$ be a smooth, compact Calabi--Yau threefold, with K\"ahler form $\omega$, a compatible hermitian metric $g_{a\bar b}$ and a nowhere-vanishing complex three-form $\Omega$. Together, $\omega$ and $\Omega$ define an $\text{SU}(3)$ structure on $X$. The statement that $g_{a\bar b}$ has $\text{SU}(3)$ holonomy is equivalent to the differential conditions
\begin{equation}
\dd\omega = 0,\qquad \dd\Omega = 0,
\end{equation}
which, in turn, imply that $X$ is Ricci-flat.
Let $x^a$, $a=1,2,3$ be the three complex coordinates on $X$. Since $g_{a \bar b}$ is hermitian, the pure holomorphic and anti-holomorphic components of the metric must vanish; that is
\begin{equation}
g_{a b}(x, \xb) = g_{\bar a \bar b}(x, \xb) = 0.
\end{equation}
Only the mixed components survive, which are given as the mixed partial derivatives of a single real scalar function, the K\"ahler potential $K$:
\begin{equation}\label{gijbar}
g_{a \bar b}(x, \xb) = \partial_a \partial_{\bar b} K(x, \xb).
\end{equation}
Note that, to simplify our notation, we will often denote the determinant of the hermitian metric by
\begin{equation}
g \equiv \det g_{a \bar b}.
\end{equation}
The K\"ahler form derived from the K\"ahler potential is
\begin{equation}
\omega = \frac{\ii}{2} \sum_{a, \bar b = 1}^3 g_{a \bar b}(x, \xb) \dd x^a \wedge \dd \bar{x}^{\bar b} = \frac{\ii}{2} \del \delb K(x, \xb),
\end{equation}
where $\del$ and $\delb$ are the Dolbeault operators. Recall that $K$ is only {\it locally} defined -- globally one needs to glue together the local patches by finding appropriate transition functions $f$ (K\"ahler transformations) so that
\begin{equation}
K(x, \xb) \sim K(x, \xb) + f(x) + \bar{f}(\xb) .
\end{equation}
Since $X$ is K\"ahler, the Ricci tensor is given by
\begin{equation}
R_{a \bar b}=\partial_{a}\partial_{\bar{b}}\ln g.
\end{equation}
Practically, finding a Ricci-flat K\"ahler metric on $X$ reduces to finding the corresponding K\"ahler potential as a real function of $x$ and $\xb$. Yau's celebrated proof~\cite{yau} of the Calabi conjecture~\cite{calabi} then guarantees that this Calabi--Yau metric is unique in each K\"ahler class.

The particle content of the low-energy theory one finds after compactifying heterotic string theory on a Calabi--Yau threefold is fixed by topological data of both the manifold $X$ and the choice of gauge bundle $V$~\cite{Candelas:1985en,Strominger:1985it,Strominger:1985ks}. The masses and couplings of the particles, roughly speaking, are then fixed by cubic couplings (with masses coming from coupling to Higgs fields). Schematically, these couplings take the form
\begin{equation}
C_{ABC}=\int_X \psi_A \cdot \psi_B \cdot \psi_C,
\end{equation}
where the $\psi_A$ are zero modes of the Dirac operator on $X$ coupled to the connection on $V$.\footnote{This is a schematic expression, since the $\psi_A$ should be contracted or wedged with $\Omega$'s so that one has a $(3,3)$-form that can be integrated over the threefold.} These zero modes have a topological origin~\cite{Strominger:1985it,Strominger:1985ks}. For the standard embedding, $V=TX$, the zero modes are related to harmonic $(1,1)$- and $(2,1)$-forms on $X$, while for more general bundles the relevant objects are $(0,p)$-forms valued in $V$ and tensor products thereof.

Note that $C_{ABC}$ does not give the \emph{physical} couplings unless the zero modes $\psi_A$ are correctly normalized.\footnote{See \cite{Douglas:2015aga} for a review of these problems.} To do this, one needs to compute the integrals
\begin{equation}
M_{A\bar B}=\int_X \omega\wedge\omega\wedge[\psi_A\cdot\bar\psi_{\bar B}],
\end{equation}
where $[\;]$ indicates a contraction with $\Omega$'s and $\bar\Omega$'s to give a $(1,1)$-form. The normalized couplings are then given by calculating $C_{ABC}$ in a basis of zero modes where $M_{A\bar B}=\delta_{A\bar B}$. Note that $M_{A\bar B}$ depends on the harmonic representative we take for the $\psi_A$ modes -- it is not enough to only know their cohomology classes. For the simplest example where $V=TX$ (and deformations thereof), one can compute $M_{A \bar B}$ using the tools of special geometry. It is not known how or if one can compute $M_{A\bar B}$ using similar tools for general choices of bundle $V$. Instead, one must tackle the problem in its full glory by finding the Ricci-flat metric on $X$, calculating the connection on $V$, and finally explicitly computing the normalized $V$-valued $(0,p)$-forms.

To date, no analytic Calabi--Yau metric has ever been found on any compact Calabi--Yau manifold (other than for trivial cases, such as products of tori). Nevertheless, an explicit algorithm to \emph{numerically} determine the Ricci-flat metric was given by Donaldson~\cite{donaldson1}. This algorithm has subsequently been explored in a variety of papers where it has been used to find numerical Calabi--Yau metrics, find gauge bundle connections that satisfy the hermitian Yang--Mills equation, examine bundle stability and explore the metric on Calabi--Yau moduli spaces~\cite{Headrick:2005ch,Douglas:2006hz,Douglas:2006rr,Doran:2007zn,Braun:2007sn,Braun:2008jp,Headrick:2009jz, Anderson:2010ke, Anderson:2011ed,Keller:2009vj}.

In the remainder of this section, we describe Donaldson's algorithm in more detail (with the specific case of the Fermat quintic presented in Appendix \ref{app:donaldson}) and discuss the computational problems one faces when trying calculate to higher order $k$ in the iterative approximation. These challenges will motivate the machine-learning approach we discuss in the remainder of the paper.

\subsection{Donaldson's algorithm}\label{sec:Donaldson_algorithm}

The general idea of Donaldson's algorithm~\cite{donaldson1} is to approximate the K\"ahler potential of the Ricci-flat metric using a finite basis of degree-$k$ polynomials $\{s_\alpha\}$ on $X$, akin to a Fourier series representation (see also~\cite{donaldson2} and \cite{tian}). This ``algebraic'' K\"ahler potential is parametrized by a hermitian matrix $h^{\alpha\bar\beta}$ with constant entries. Different choices of $h^{\alpha\bar\beta}$ correspond to different metrics within the same K\"ahler class. Following Donaldson's algorithm, one then iteratively adjusts the entries of $h^{\alpha\bar\beta}$ to find the ``best'' degree-$k$ approximation to the unique Ricci-flat metric. Here ``best'' is taken to mean the \emph{balanced} metric at degree $k$. Note that, as one increases $k$, the balanced metric becomes closer to Ricci-flat at the cost of exponentially increasing the size of the polynomial basis $\{s_\alpha\}$ and the matrix $h^{\alpha\bar\beta}$. As we will discuss, at some point it becomes computationally extremely difficult to further increase $k$ and, hence, to obtain a more accurate approximation to the Ricci-flat metric..

One can check how good the approximation is -- that is, how close the balanced metric is to being Ricci-flat for a given value of k -- by computing a variety of ``error measures''. These include $\sigma$, a measure of how well the Monge--Amp\`ere equation is solved, and $\Vert R\Vert$, a direct measure of how close the metric is to Ricci-flat. We will describe exactly what these quantities are later in this section. 

Let us begin by summarizing the algorithm as given by Donaldson. After this we will discuss how one implements it numerically, with many of the details left to Appendix \ref{app:donaldson}.
\begin{enumerate}
	
	\item Let the degree $k$ be a fixed positive integer. We denote by $\{s_\alpha \}$ a basis of global sections\footnote{More generally, one takes sections of an ample line bundle over $X$~\cite{Braun:2007sn}.}  of $\cO_X(k)$:
	\begin{equation}
	H^0(X, \cO_X(k)) = \operatorname{span} \{ s_\alpha \} , \qquad \alpha = 1, \ldots, N_k.
	\end{equation}
	In other words, we choose an $N_k$-dimensional basis of degree-$k$ holomorphic polynomials $s_\alpha(x)$ on $X$. The values of $N_k$ grow factorially with $k$; for a quintic Calabi--Yau, $N_k$ is given for any $k$ by equation \eqref{Nk}.
	
	\item Make an ansatz for the K\"ahler potential of the form
	\begin{equation}\label{eq:KP}
	K(x, \xb) = \frac{1}{k \pi} \ln \sum_{\alpha, \bar{\beta} = 1}^{N_k} h^{\alpha \bb} s_{\alpha}(x) \sb_{\bb}(\xb),  
	\end{equation} 
	where $h^{\alpha \bb}$ is some invertible hermitian matrix. As we show in Appendix \ref{sec:metric_sigma}, one can use expression \eqref{gijbar} to obtain the components of the corresponding metric $g^{(k)}_{a \bar b}$ from this expression for $K$.
	
	\item The pairing $h^{\alpha \bb} s_{\alpha} \sb_{\bb}$ defines a natural inner product on the space of global sections, so that $\hmat$ gives a metric on $\cO_X(k)$. Consider the hermitian matrix
	\begin{equation}\label{Hh}
	H_{\alpha \bb} \equiv  \frac{N_k}{\vcy{X}} \int_X \dd \vcy{X}
	\frac{s_\alpha \sb_{\bb}}{ h^{\gamma \db} s_{\gamma} \sb_{\db}},
	\end{equation}
	where $\vcy X$ is the integrated volume measure of $X$:
	\begin{equation}
	\dd \vcy{X} = \Omega \wedge \bar{\Omega}.
	\end{equation}
	In general, $\hmat$ and $H_{\alpha\bar\beta}$ will be unrelated. However, if they are inverses of each other
	\begin{equation}\label{balanced}
	h^{\alpha \bb} = ( H_{\alpha \bb} )^{-1} ,
	\end{equation}
	the metric on $\cO_X(k)$ given by $\hmat$ is said to be ``balanced''. This balanced metric then defines a metric $g^{(k)}_{a \bar b}$ on $X$ via the K\"ahler potential \eqref{eq:KP}. We also refer to this metric on $X$ as balanced.
	
	\item Donaldson's theorem then states that for each $k \geq 1$ a balanced metric exists and is \emph{unique}. Moreover, as $k\to\infty$, the sequence of metrics $g^{(k)}_{a \bar b} = \partial_a \partial_{\bar b} K$ converges to the unique Ricci-flat K\"ahler (Calabi--Yau) metric on $X$.
	
	\item In principle, for each $k$, one could solve \eqref{balanced} for the $\hmat$ that gives the balanced metric using \eqref{Hh} as an integral equation. However, due to the highly non-linear nature of the equation, an analytic solution is not possible. Fortunately, for each integer $k$, one can solve for $\hmat$ iteratively as follows:
	\begin{enumerate}
		\item Define Donaldson's ``$T$-operator'' as
		\begin{equation}\label{eq:T_operator}
		T \colon h^{\alpha \bb}_{(n)} \mapsto   T(h_{(n)})_{\alpha \bar{\beta}} = \frac{N_k}{\vcy{X}} \int_X \dd \vcy{X}
		\frac{s_\alpha \sb_{\bb}}{ h^{\gamma \db}_{(n)} s_{\gamma} \sb_{\db}} .
		\end{equation}
		\item Let $h^{\alpha\bar\beta}_{(0)}$ be an initial invertible hermitian matrix.
		\item Then, starting with $h^{\alpha\bar\beta}_{(0)}$, the sequence
		\begin{equation}
		h_{(n+1)} = [ T(h_{(n)}) ]^{-1}
		\end{equation}
		converges to the desired balanced metric $\hmat$ as $n \to \infty$.
	\end{enumerate}
	The convergence is very fast in practice, with only a few iterations ($\lesssim 10$) necessary to give a good approximation to the balanced metric. For all calculations in this paper, we {\it iterate the $T$-operator ten times}.
\end{enumerate}

At this point, one has an approximation to the Calabi--Yau metric, given by the balanced metric $g^{(k)}_{a \bar b} = \partial_a \partial_{\bar b} K$ computed at degree $k$. A natural question is: just how good is this approximation, that is, how close is the balanced metric evaluated for integer $k$ to being Ricci-flat? A number of ``error measures'' have been introduced in the literature for this purpose~\cite{Douglas:2006rr,Douglas:2006hz,Braun:2007sn,Anderson:2010ke}, two of which we discuss here.
\paragraph*{$\sigma$ measure:} The ``$\sigma$ measure'' is a measure of Ricci flatness encoded by the Monge--Amp\`ere equation. Consider the top-form $\omega^3$ defined by the K\"ahler form. Since $X$ is a Calabi--Yau threefold with $\Omega  \wedge \overline{\Omega}$ the unique (up to scaling) non-vanishing $(3,3)$-form, these two must be related by an overall constant $c$. That is
	\begin{equation}
	\omega\wedge\omega\wedge\omega = c \, \Omega  \wedge\Ob .
	\end{equation}
	This is equivalent to the Monge--Amp\`ere equation which defines the Calabi--Yau metric. Comparing $\Omega\wedge\bar\Omega$ with $\omega^3$, one should find they agree \emph{pointwise} up to an overall constant $c$ (which is the same for all points). To avoid computing the constant, we can compare the integral of the two top-forms so that $c$ cancels:
	\begin{equation}
	\begin{array}{l}
	\vk{X} = \int_X \omega^3 \\
	\vcy{X} = \int_X \Omega  \wedge \Ob 
	\end{array}
	\qquad \Rightarrow \qquad
	\frac{\omega^3}{\vk{X}} = \frac{ \Omega  \wedge \Ob}{\vcy{X}} .
	\end{equation} 
	Note that one can compute $\Omega$ exactly using a residue theorem. Taking $\omega=\omega_k$, where $\omega_k$ is the K\"ahler form for $g^{(k)}_{a \bar b}$, this equality holds if and only if $\gk$ is the desired Calabi--Yau metric. Said differently, the ratio of $\omega^3_k/\vk{X}$ and $\Omega \wedge \overline{\Omega}/\vcy{X}$ must be 1. Integrating over $X$, the quantity
	\begin{equation}\label{eq:sigma_int}
	\sigma_k \equiv \frac{1}{\vcy{X}} \int_X \dd {\vcy{X}}\, \left| 
	1 - \frac{\omega^3_k \slash \vk{X}}{\Omega  \wedge \overline{\Omega} \slash \vcy{X} }
	\right|
	\end{equation}
	is 0 if and only if $\omega_k$ is the K\"ahler form of the Calabi--Yau metric. In other words, $\sigma_k$ is a measure of how far $\gk$ is from the Ricci-flat metric. As $k$ increases, $\sigma_k$ approaches zero at least as fast as $k^{-2}$~\cite{Braun:2007sn,Braun:2008jp}. Note that $\omega^3_k$, and thus $\sigma_k$, can be computed directly from the determinant of the metric $\detg k \equiv \det g^{(k)}_{a \bar b}$. This is one of the reasons we focus on the determinant later in this paper: it is straightforward to check how accurate our machine-learning approach is by computing $\sigma_k$.
		
\paragraph*{$\Vert R \Vert$ measure:} The ``$\Vert R \Vert$ measure'' is a global measure of how close to zero the Ricci scalar is. The quantity
	\begin{equation}\label{eq:R_int}
	\Vert R \Vert_k \equiv \frac{\vk{X}^{1/3}}{\vcy{X}} \int_X \dd {\vk{X}}\,|R_k|,
	\end{equation}
	where $R_k$ is the Ricci scalar computed using the balanced metric for integer $k$, is zero if and only if $\gk$ is the exact Calabi--Yau metric. The various factors of $\vk{X}$ and $\vcy{X}$ that appear in this expression are there to remove any scaling dependence on $k$.\footnote{This is discussed in more detail in \cite{Anderson:2010ke}.} As $k$ increases, $\Vert R \Vert_k$ tends to zero as $k^{-1}$~\cite{Anderson:2010ke}.
		
Note that there are other error measures one could use, such as the $\Vert EH\Vert$ measure from \cite{Anderson:2010ke}, or the pointwise values of the Ricci scalar $R$ or any components of the Ricci tensor $R_{a\bar b}$. However, we will not discuss them in this paper. We leave the details of the numerical implementation of Donaldson's algorithm and the calculation of the error measures $\sigma_k$ and $\Vert R \Vert_k$  to Appendices A and B. They are also discussed in several reviews in the literature~\cite{Douglas:2006rr,Douglas:2006hz,Braun:2007sn,Anderson:2010ke,Anderson:2011ed}. Here we focus only on those details which will be relevant to machine-learning the Calabi--Yau metric later in this paper.

Both the $T$-operator and the error measures involve integrating over the threefold, suggesting that we need to introduce local coordinate charts and all of the complications that come with these. Fortunately, we can avoid this by approximating integrals by sums over random\footnote{There are nuances concerning which random distribution to use and how this effects the integration measure~\cite{Douglas:2006rr}. We comment on this in more detail in the discussion surrounding \eqref{w} in Appendix A.} points ${p_M}$ on $X$:
\begin{equation}
\int \text{dVol}\, f \sim \frac{1}{N}\sum_{M=1}^{N} f(p_M).
\end{equation}
The number of points we need to take when approximating the integrals is important and will be explicitly discussed. In addition, once we have found the balanced metric $\hmat$ for fixed integer $k$, we need to consider how many points at which to evaluate $\gk$. That is, there are three (unrelated) numbers of points that must be specified. These are:
\begin{itemize}
	\item Let $N_p$ be the number of random points we sum over to approximate the $T$-operator in \eqref{eq:T_operator}. As we discuss below, once we have chosen the degree $k$ -- and, hence, $N_{k}$ -- at which to approximate the Ricci-flat metric, $N_p$ is bounded by the requirement that $N_p\gg N_k^2$.
	\item Let $N_t$ be the number of test points we use to compute the error measures $\sigma_k$ and $\Vert R\Vert_k$ in \eqref{eq:sigma_int} and \eqref{eq:R_int}.
	\item Let $N_g$ be the number of points for which we want to know the value of the metric $\gk$.
\end{itemize}
We consider each of these in turn.

As discussed in \cite{Braun:2007sn}, since the $T$-operator leads to an $N_k\times N_k$-matrix $\hmat$, one needs $N_p \gg N_k^2$ points for convergence of $\hmat$ to the balanced metric. If one uses too few points, one finds $\hmat$ does not converge properly to the balanced representative and the resulting metric $\gk$ is further away from Ricci-flat than one would otherwise expect. Said differently, iterating the $T$-operator with too few points leads to an $\hmat$-matrix, and the associated metric on $X$, that has error measures larger than those of the $\hmat$ matrix computed with $N_p \gg N_k^2$. It was found in a previous study of the scalar Laplacian~\cite{Braun:2008jp} that a good choice is
\begin{equation}\label{eq:point_bound}
N_p =10\,N_k^2 + 50{,}000.
\end{equation}
Unless otherwise stated, for fixed $k$ and hence $N_{k}$, we will always evaluate the $T$-operator using this many points $N_{p}$ in the integral.

We do not need to evaluate the integrals in the error measures for all $N_p$ points. Instead, one can approximate these error measure integrals, for any integer $k$, with a fixed number of points $N_t$. Roughly, the percentage error when computing, for example, $\sigma$ with $N_t$ points is $N_t^{-1/2}$. Hence, if one is interested only in checking how close the numerical metric is to Ricci-flat, it is sufficient to take $N_t=10{,}000$ to get estimates that are good to 1\%. Of course, using larger values for $N_{t}$ will result in even more accurate results. In the remainder of this paper, we will explicitly state the value of $N_{t}$ that we are choosing for a given calculation.

Finally, for a fixed value of $k$, $N_g$ is the number of points for which we want to know the value of the components of the resulting metric $\gk$. This is the desired output of Donaldson's algorithm, giving us the ability to calculate the metric numerically and then to use it to compute other quantities on the Calabi--Yau threefold. For example, if one wants to solve the hermitian Yang--Mills equations~\cite{Douglas:2006hz,Anderson:2010ke,Anderson:2011ed} or find the eigenmodes of Laplace operators~\cite{Braun:2008jp}, one needs to know $\gk$ numerically. In the latter case, it was found in ~\cite{Braun:2008jp} that for a quintic Calabi--Yau threefold it is sufficient to solve for the eigenmodes using 500,000 random points. Since the metric appears in the Laplace operator, one also needs to know $\gk$ for those same random points; that is, $N_g=500{,}000$. Similarly, when we predict the values of $\detg k$ on the quintic later in this paper, we will assume that we want to know these values for $N_g=500{,}000$ random points.

Note that the bound \eqref{eq:point_bound} has implications for both the speed and feasibility of the numerical calculations. For example, for a quintic hypersurface embedded in $\mathbb{P}^4$, using \eqref{Nk} for $k=5$ and $k=10$ one has $N_k=125$ and $N_k=875$ respectively. This means one needs $N_p=206{,}250$ and $N_p=7{,}706{,}250$ points respectively to be confident that the $T$-operator will converge properly to the balanced metric. For $k=20$, $N_p$ needs to be on the order of 500 million points. We see that as we push to higher values of $k$, in addition to the size $N_k$ of the polynomial basis $\{s_\alpha\}$ increasing factorially, the number of points we sum over to find the balanced metric also grows factorially. Together, these make going to larger values of $k$ prohibitive in both time and computational resources. Previous studies have been limited to values of $k$ that give on the order of 500 sections (roughly $k=8$ or so for the Fermat quintic).

\subsection{A check using Donaldson's algorithm}

\begin{figure}[!!!h]
	\centerline{
		\hspace{-3.25em}\includegraphics[trim=0mm 0mm 0mm 0mm, clip, width=10cm]{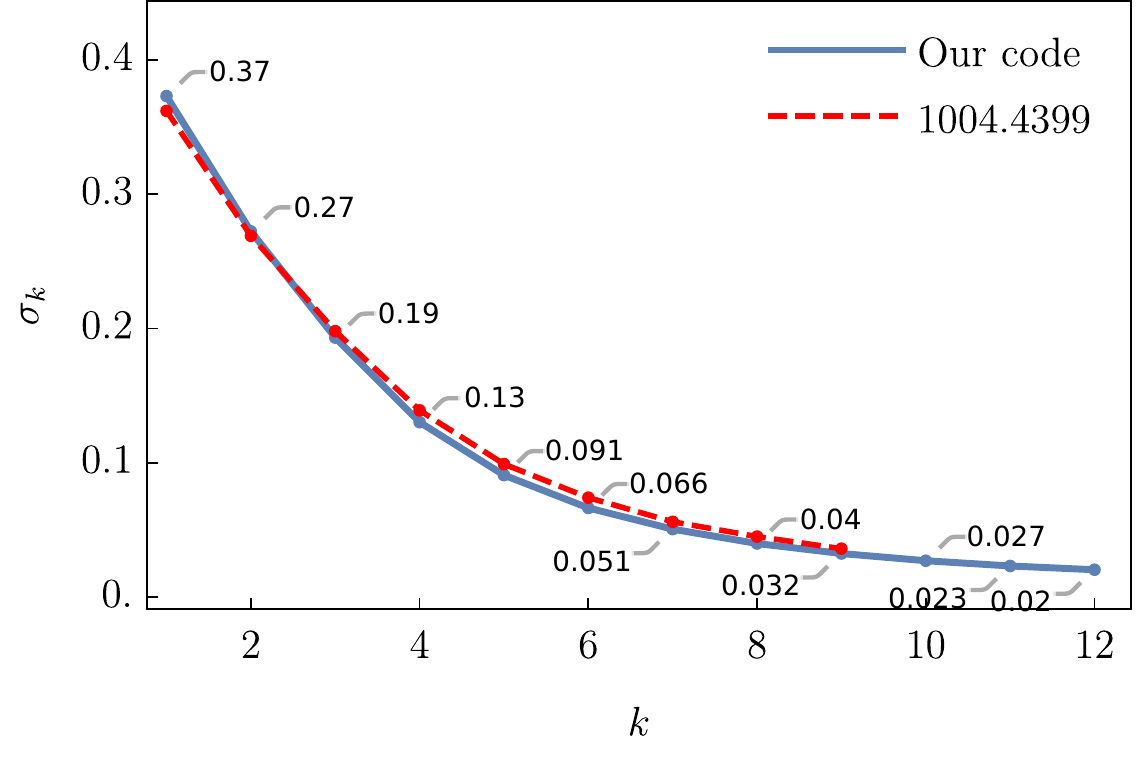}
	}
	\caption{Behavior of the $\sigma$ error measure as $k$ increases. This was computed using Donaldson's algorithm with $N_p=10\,N_k^2+50{,}000$ points for the $T$-operator integral and $N_t=5\times10^5$ test points for the $\sigma$ integral. The blue line is computed using our Mathematica implementation, while the dashed red line corresponds to previous results from reference 1004.4399~\cite{Anderson:2010ke}. They are in close agreement.
		\label{fig:sigma_Donaldson}}
\end{figure}

\begin{figure}[!!!h]
	\centerline{
		\hspace{-3.25em}\includegraphics[trim=0mm 0mm 0mm 0mm, clip, width=10cm]{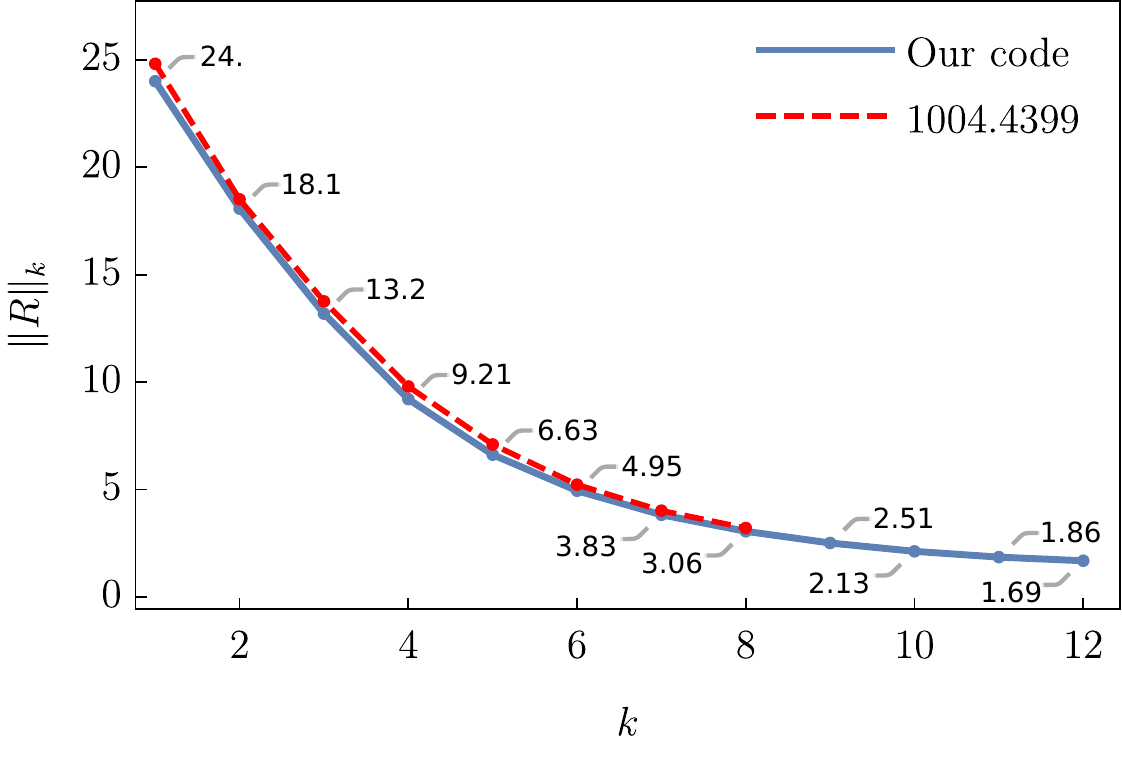}
	}
	\caption{Behaviour of the curvature measure $\Vert R \Vert$ measure as $k$ increases. This was computed using Donaldson's algorithm with $N_p=10\,N_k^2+50{,}000$ points for the $T$-operator integral and $N_t=5\times10^5$ test points for the $\Vert R \Vert$ integral. The blue line is computed using our Mathematica implementation, while the dashed red line corresponds to previous results from reference 1004.4399~\cite{Anderson:2010ke}. They are in close agreement.
		\label{fig:R_abs_Donaldson}}
\end{figure}


Although all analytic and numerical methods that we will discuss are valid on any threefold, for concreteness we focus on the Fermat quintic $Q$, defined by equation \eqref{defQ}, for the remainder of the paper.  The main aim of this paper is to use machine learning to enhance the speed of calculation of the Calabi--Yau metric, reducing the time needed compared to using Donaldson's iterative algorithm.  As we discuss in Appendix \ref{app:numerical}, we have chosen to implement this using Mathematica, rather than C\texttt{++} as was previously used, since it is well suited to the numerical linear algebra calculations that occur in Donaldson's algorithm and provides an extensive suite of machine-learning tools. 
As a check of our Mathematica implementation, we first apply it explicitly to Donaldson's algorithm, compute the various error measures, and compare these with the error measures previously found in \cite{Anderson:2010ke} using a C\texttt{++} implementation. In Figures \ref{fig:sigma_Donaldson} and \ref{fig:R_abs_Donaldson}, we plot $\sigma_k$ and $\Vert R \Vert_k$ respectively for $k=1,\ldots,12$, for both our new Mathematica implementation -- the blue line -- and the C\texttt{++}  implementation -- the red dashed line -- used in \cite{Anderson:2010ke}. In all cases, calculations of the $T$-operator were carried out using $N_p$ points, fixed by \eqref{eq:point_bound}. The error measures were computed using $N_t=\text{500,000}$ for all $k$. We use $N_t=\text{500,000}$ here so that we can directly compare our Mathematica implementation with the results in \cite{Anderson:2010ke} which were computed using $N_t=\text{500,000}$. We conclude that the C\texttt{++} results are reproduced by our new Mathematica implementation, which we employ from here onwards.

\begin{figure}[!h!t!b]
	\centerline{
		\hspace{-3.25em}\includegraphics[trim=0mm 0mm 0mm 0mm, clip, width=10cm]{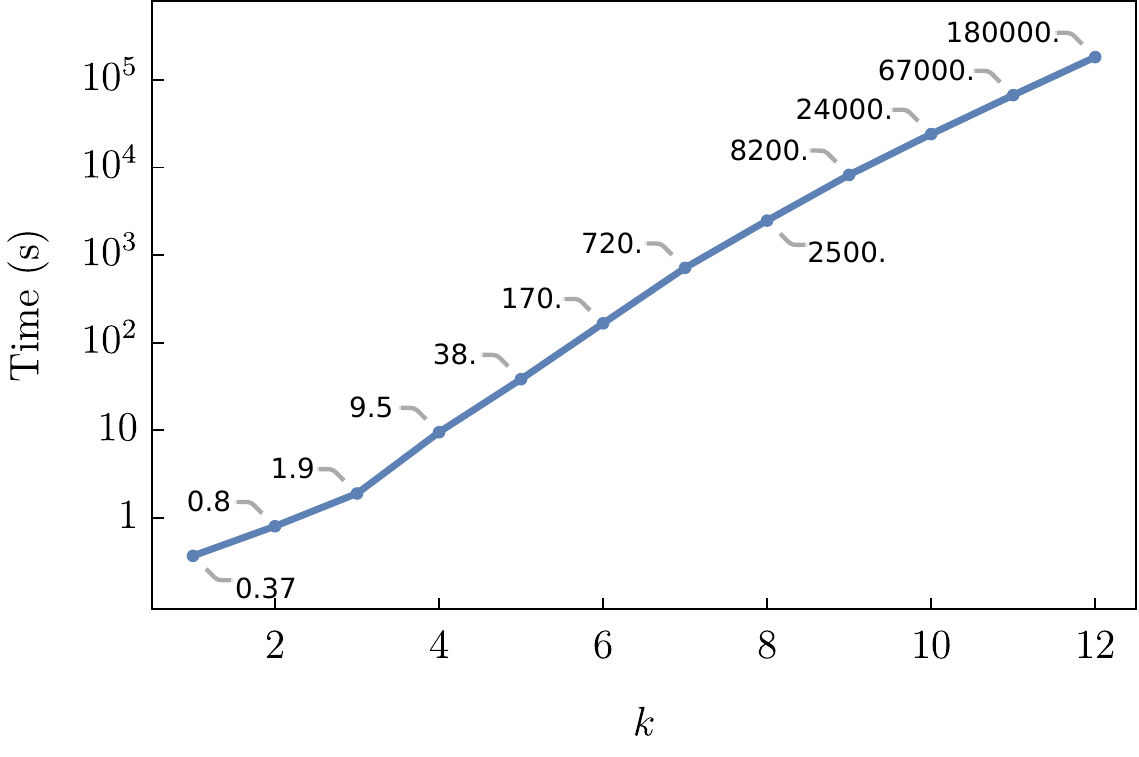}
	}
	\caption{Plot of the time taken in seconds for our Mathematica programming to find the balanced $\hmat$-matrix as $k$ varies from 1 to 12. This was computed using Donaldson's algorithm with $N_p=10\,N_k^2+50{,}000$ points for the $T$-operator integral, where $N_k$ is given by \eqref{Nk}. Times are given to two significant figures.}
		\label{fig:h_time}
\end{figure}

Note that the numerical approximation to the Ricci-flat metric improves as $k$ increases. Unfortunately, as $k$ increases, the computational time and resources needed to carry out the numerical integrations grow dramatically. This is due to factorial growth of both the number of sections $N_k$ and the number of integration points $N_p$. In Figure \ref{fig:h_time}, we plot the times needed to calculate the $\hmat$-matrix as $k$ varies. We do indeed see factorial growth. For $k=12$, calculations take on the order of 50 hours. Such long times might be acceptable if one is interested only in computing the balanced metric once to high accuracy. However, in reality, one would like to vary the complex structure or K\"ahler parameters to explore the moduli space of the Calabi--Yau threefold without reducing the accuracy of the approximation. In the case of gauge connections, one would like to employ similar methods to explore gauge bundle stability. For both of these, one needs to repeatedly calculate the $\hmat$-matrix quickly -- a single calculation that takes 50 hours suddenly looks very slow when one has to repeat it for multiple choices of complex structure, K\"ahler and bundle moduli. Note that it is unlikely that one will be able to go to much larger $k$ values in the near future using Donaldson's algorithm alone without moving to a cluster -- for example, the above timings would suggest that $k=20$ would take approximately 35 years!

Ideally, we would like to find some way of greatly speeding up this calculation and improving the accuracy of our approximation (akin to going to larger values of $k$). It is clear that, to do so, one must modify the calculational procedure and no longer use Donaldson's algorithm on its own. The remainder of this paper discusses how this might be done using a combination of machine learning and curve fitting.

\section{Machine learning the Calabi--Yau metric}
In \cite{He:2017aed,He:2017set} a paradigm was proposed to use artificial intelligence in the form of machine learning (deep learning in particular) to bypass expensive algorithms in computational geometry.  
Indeed, \cite{He:2017aed,He:2017set,Krefl:2017yox,Ruehle:2017mzq,Carifio:2017bov} brought about much collaboration between the machine-learning and string theory communities.
It was found that problems central to formal and phenomenological aspects of string theory, such as computing bundle cohomologies or determining particle spectra, appear to be machine learnable to very high precision (see \cite{He:2018jtw} for a pedagogical introduction).
It is therefore natural to ask whether machine-learning techniques may be of use in our present, computationally expensive problem.
{\it Henceforth, we will abbreviate any machine-learning techniques, be they neural networks or decision trees, collectively as ML.}

As one can see from \eqref{Nk}, the size of the monomial basis $\{s_\alpha\}$ at degree $k$, and hence the size of the matrix $\hmat$, grows factorially. This presents a problem: the Ricci-flat metric is better approximated as $k\to\infty$, but the complexity growth with respect to $k$ is factorial. Furthermore, Donaldson's algorithm involves evaluating the monomial basis at each sampled point $p$ on the manifold, multiple matrix multiplications and finally a matrix inverse. Taken together, it is clear that pushing the algorithm to higher values of $k$ is, at best, computationally expensive and, at worst, impossible with reasonable bounds on accessible hardware.

If one could \emph{predict} the relevant quantities at higher $k$ given data computed at lower $k$, then one could \emph{bypass} the most expensive steps of Donaldson's algorithm. In this section, we discuss how this can be done given a small sample of values at the higher value of $k$. Note, however, that the small sample of values must still be computed by following Donaldson's algorithm -- we still need to evaluate the $T$-operator for $N_p\gg N_k^2$ points to find the balanced metric. This section should therefore be seen as a \emph{test} of our machine-learning approach. However, to be useful in practice, we must find some way of calculating or predicting the higher $k$ values \emph{without}  Donaldson's algorithm; in the later sections, we outline how this can be done.

\subsection{Supervised learning}

We begin with a somewhat abstract review of machine learning, focussing on the particular case of supervised learning. We will try to make this more concrete in Section \ref{sec:learn_det}, where we show how this applies to the problem at hand.

Our problem is a natural candidate for \emph{supervised learning}:
\begin{itemize}
\item We have a set of input values $I_i$ for which {\it we know} the output $O_i$. For example, the inputs might be a set of points on the quintic and the values of $\detg 1$ at each such point. The outputs might be the values of $\gk$ or $\detg k$ at each point for a larger value of $k$. This constitutes a set of {\em labelled data} $\mathcal{D}$ of the form $\mathcal{D}=\{I_i\} \to \{O_i\}$.
\item Using this data, we can train an appropriate ML (which can be any of the standard ones such as a neural network, a classifier, a regressor, etc.) to predict output values $\hat{O}_i$ from the inputs $I_i$.\footnote{We will denote predicted quantities with hats throughout the paper.} Here, training means that the ML optimizes its parameters in order to minimize some cost-function (such as the mean squared error, determined by how far off the predicted values $\hat{O}_i$ are from the actual values of $O_i$).
\item Given a set of new inputs $I_j$ for which we do not know the outputs, we use the trained ML to predict a set of outputs, one $\hat{O}_j$ for each $I_j$.
\end{itemize}
In its simplest form, supervised learning is no different from regression, familiar from rudimentary statistics. The key difference with supervised learning (and machine learning more generally) is that one does not specify a single, usually quite simple, function as in the case of regression, but rather a set of non-linear functions, such as a complicated directed graph of nodes in the case of neural networks, or multi-level output in the case of classifiers. The more sophisticated the structure of the ML, the better it can approximate complicated systems.

In general, one needs a measure of how well trained the ML is; that is how accurate its predictions are. The standard measure uses \emph{cross validation}. Take the labelled data $\cD = \{I_i\} \to \{O_i\}$ and split it into two complementary sets, $\cT$ and $\cV$, so that $\cD = \cT \cup \cV$. These are usually referred to as the training data, $\cT$, and the validation data, $\cV$. Cross-validation is as follows:
\begin{itemize}
\item We train the ML on $\cT$, the training set. This optimises the parameters of the ML to minimise whichever cost-function we pick.
\item We apply the optimized ML on the inputs $I_j$ of $\cV$, the validation set, giving us a set of predicted values $\hat{O}_j$.
\item We then cross-check the predicted values $\hat{O}_j$ against the known values $O_j$ within the validation set $\cV$. We do this by examining some goodness-of-fit measure $G$ (such as percentage agreement or chi-squared). This allows us to see how well the ML is performing.
\item We then vary the size of the training set $\cT$ to see how the goodness-of-fit measure $G$ varies. For example, we could check how well the ML performs after training on 10\%, 20\%, etc., of the total data $\cD$. The plot of $G$ against the size of $\cT$ is called the \emph{learning curve}. Typically, the learning curve is a concave function that increases monotonically as we increase the percentage of training data. In other words, when the ML is trained on a larger sample of data, it performs better, but the improvement diminishes with each added training point.
\end{itemize}
The particular flavour of ML that we have chosen to focus on is that of \emph{gradient-boosted decision trees}.\footnote{
	We comment here that we have tried some other ML structures, such as the forward-feeding multi-layer perceptron neural network which was shown to be very well adapted to computing cohomology~\cite{He:2017aed,He:2017set}. Interestingly, these do not seem to perform any better.}  The details of this are not important for what follows -- we give a overview of this particular approach in Appendix \ref{sec:gbt}. We now discuss how supervised learning applies to the problem of Ricci-flat K\"ahler metrics.

\subsection{Learning the determinant}\label{sec:learn_det}

One expects the analytic form of the K\"ahler potential $K$ for a Calabi--Yau metric to be a complicated non-holomorphic function; so complicated, in fact, that no explicit form has ever been written down, even for the simplest of compact Calabi--Yau manifolds.\footnote{Excluding the $n$-torus and products thereof.} This is why numerical metrics are the best one can do for now.

As we discussed in Section \ref{sec:Donaldson_algorithm}, Donaldson's algorithm gives a way to approximate the honest Ricci-flat metric on $Q$ via a balanced metric $g^{(k)}_{a \bar b}$, computed at some fixed degree $k$. Since the metric, its determinant and the Ricci tensor can be derived in turn by simple operations such as logarithms and derivatives, we choose to focus on one of them. 

Let us consider the determinant of the metric, $\detg k$, because:
\begin{enumerate}
	\item It is a convenient scalar quantity, easily calculated from $\gk$ itself.
	\item It encodes curvature information since the mixed partials of its logarithm give the Ricci tensor.
	\item It allows one to integrate quantities over the manifold.
	\item One can use it to compute the accuracy measure $\sigma_k$. From \eqref{eq:sigma_int}, the only approximate quantity appearing in $\sigma_k$ is $\omega_k^3$, but this is fixed by the determinant of $g^{(k)}_{a \bar b}$ via the relation
	\begin{equation}
	\omega^3_k \propto \detg k\, \dd x^1\wedge\dd\bar{x}^1\wedge\ldots\wedge\dd x^3\wedge\dd\bar{x}^3.
	\end{equation}
\end{enumerate}
In other words, $\detg k$ gives a straightforward example to which we can apply machine-learning techniques while still allowing us to compute the $\sigma_k$ error measure to check the accuracy of our methods. We could, for example, have focussed on the K\"ahler potential itself, but since we are predicting the \emph{values} at each point and not its functional form, we would have been unable to compute $\sigma_k$ to check whether our approach was actually useful.

Since the metric itself can be thought of as a collection of patch-wise functions $g_{a \bar b}(x,\bar x)$, our procedure for predicting $\detg k$ can also be used to predict the values of $\gk$ itself. This is, of course, what we actually want to do in practice since it is the metric itself that enters into calculations of the gauge connection and various Laplace operators on $Q$.\footnote{Moreover, since one can extract the exact Calabi--Yau determinant from $\Omega\wedge\bar\Omega$, the values of $\detg k$ are not interesting in their own right. One should think of this paper as giving a prescription that also applies to other quantities, such as the components of the metric or gauge connection.} For the moment however, for simplicity, let us concentrate on the determinant and remember that everything we say can easily be applied to the metric itself.

Given the supervised learning routine outlined above, it is natural to ask whether machine-learning techniques can improve the accuracy of our approximation to the Calabi--Yau metric and/or reduce the amount of time needed for the calculation. Specifically, focussing on the determinant, we ask:
\begin{quote}
Given a set of points on the quintic $Q$ and the corresponding values of $\detg{l} $ computed at some low degree $l$, can one \emph{predict} the values of $\detg k$ computed at a higher degree $k>l$?
\end{quote}
As an example, imagine we want the value of the Calabi--Yau determinant $g$ for $N_g=500{,}000$ points on the quintic. Using Donaldson's algorithm, we can find an approximation to $g$ by computing the determinant $\detg k$ of the balanced metric, where the degree $k$ controls the accuracy of the approximation. As we increase $k$, we get a better approximation to the honest Calabi--Yau determinant with the price being an explosion in computational time due to the factorial increases in both $N_p$ and $N_k$.\footnote{Recall that $N_k$ is the size of the monomial basis $\{s_\alpha\}$ and  $N_p$ is the number of points used in the iteration of the $T$-operator. Note that $N_p$ is unrelated to the 500,000 points at which we want to compute the value of $\detg k$. Furthermore, thanks to \eqref{eq:point_bound}, one needs to take $N_p$ to be large (and greater than 500,000) even for relatively small values of $k$.} There are then two different but related questions:
\begin{enumerate}
	\item Suppose we use Donaldson's algorithm to compute the value of $\detg l$ for \emph{all} of the $N_{g}=500{,}000$ points and $\detg k$, for fixed $k>l$, for only a small sample of them  -- can we use machine learning to predict the remaining values of the determinant $\detg k$?
	\item Suppose we use Donaldson's algorithm to compute the value of $\detg l$ for \emph{all} of the $N_{g}=500{,}000$ points and $\detg k$, for fixed $k>l$, for \emph{none} of them  -- can we use machine learning to predict all of the values of $\detg k$?
\end{enumerate}
In the first case, one needs some of the values of $\detg k$ as an input to our supervised-learning model, while, in the second, one does not need to have calculated $\detg k$ at all. The first problem is amenable to supervised learning since we have both input ($\detg l$) and output ($\detg k$) data. The second problem, unlike the first, does not naturally fall within supervised learning. Obviously, we would like to find a solution to the second problem as it would side-step having to follow Donaldson's algorithm for the higher value of $k$, whereas in the first case we still need to compute some of the higher $k$ data. 

Let us make clear why computing even a small sample of the higher $k$ data is unacceptable in practice. In the first problem, we need the value of $\detg k$ for a small sample of the $N_{g}=500{,}000$ points so that we have some data for our ML to learn from. In order to compute \emph{any} of these values, however, one must first iteratively solve for $\hmat$ at degree $k$. But we cannot simply compute the $\hmat$ matrix using only the small sample of points we intend to use as the input data! Instead, \eqref{eq:point_bound} forces us to integrate over sufficient points so that $N_p\gg N_k^2$ holds. This means there is a hidden, and unacceptably large, computational cost in solving the first type of problem.

For example, imagine we tried to teach an ML to predict $\detg {8}$ from $\detg 1$. To generate the $\detg {8}$ values in the first place, we would have to evaluate the $T$-operator for approximately $N_p=\text{8,000,000}$ points, otherwise $\hmat$ would not converge properly to the balanced metric.\footnote{For $k=8$, one finds $10\,N_k^2+50{,}000=7{,}706{,}250$.} This means that a solution to the first problem does not avoid Donaldson's algorithm (the calculation of $\hmat$ for $k=8$). Instead it avoids only the calculation of the \emph{values} of $\detg {8}$ from $\hmat$. It is an interesting {\it proof of principle} to see that machine learning can indeed learn how to approximate such complicated functions, but this clearly is {\it not all that helpful} in practice.

A solution to the second type of problem, where we do not require any of the higher $k$ data, would indeed avoid Donaldson's algorithm for the higher value of $k$ and so potentially greatly speed up the time of calculation -- this is the main goal of the paper. We will spend the remainder of this section discussing the first problem, leaving a solution to the second problem to Sections \ref{sec:curve_extrap} and \ref{sec:learn_extrap}.

Let the input values be of the form
\begin{equation}
I = \{
p = 
(z_0, z_1, z_2, z_3, z_4)
,
 \detg 1|_p
\},
\end{equation}
where $z_i$ are coordinates\footnote{For each point, one of the $z_i$ is equal to 1, corresponding to the affine patch of choice.} on $\mathbb{P}^4$ for each of $N_{g}=\text{500,000}$ random points $p\in Q$, and $\detg 1 |_p$ is the value of the determinant at the point $p$ calculated using the balanced metric at $k=1$.\footnote{In practice, because machine-learning algorithms usually take real inputs, we split each $z_i$ into real and imaginary parts. Since $g_{a \bar b}$ is a hermitian matrix, $g$ is always real. Thus, the inputs are real 11-tuples and the outputs are real numbers.} 

Next, let the output be 
\begin{equation}
O=\{\detg k |_p\},
\end{equation}
where $k>1$. As $k$ increases Donaldson's algorithm becomes more costly, both in time and computational resources, due to the sizes of the intermediate matrices involved. The idea is to avoid this by training a model to predict the values of the determinant.

To summarize, we have labelled data of the form
\begin{equation}\label{detgData}
\cD_{1,k}=\{p , \detg 1|_p \}
\to \{\detg{k}|_p \},
\end{equation}
where $\mathcal{D}_{1,k}$ signifies a data set with the values of $\detg 1$ as inputs and $\detg k$ as outputs.  Using this data structure, we will perform supervised learning and see whether an ML can accurately predict the determinant for $k>1$ from the $k=1$ values. We will then explore how the accuracy changes as we use higher values of $k$ as an input.

\subsection{Warm-up: \texorpdfstring{$k=1$}{k=1} to \texorpdfstring{$k=2$}{k=2}}

As a warm-up, let us first try learning the values of $\detg 2|_p$ from the points and $\detg 1|_p$. Donaldson's algorithm and the calculation of $\detg 2$ at $k=2$ are relatively fast, so it is easy to check how well the machine-learning model is doing.

Note that, even though we are eventually interested in the values of $\detg 2|_p$ at all $N_g=\text{500,000}$ random points, it is sufficient to limit the data to a much smaller set of such points when checking the validity of our machine-learning algorithm. 
Here, we take our labelled data $\cD_{1,2}$ to consist of $N_g=\text{10,000}$ random points on the quintic, together with the values of the determinants of the balanced metrics computed by Donaldson's algorithm at $k=1$ and $k=2$. These are organised as in \eqref{detgData}:
\begin{equation}
\cD_{1,2}=\{p , \detg 1|_p \}
\to \{\detg{2}|_p \}.
\end{equation}
In principle, there is some complicated function which describes this map. Standard regression analysis would require one to guess some non-linear function with parameters which approximates this map, and then optimise the parameters using least-squares, etc. However, even the form of this function is difficult to imagine. Herein lies the power of machine-learning: one does not try to fit a single function, but rather, uses a combination of non-linear functions or decision trees in an interactive and interconnected fashion. The ML can then, in principle, approximate the function without us having to guess its form in the first place.

\begin{figure}[!h!t!b]
	\centering
	\begin{subfigure}{.5\textwidth}
		\centering
		\includegraphics[width=7cm]{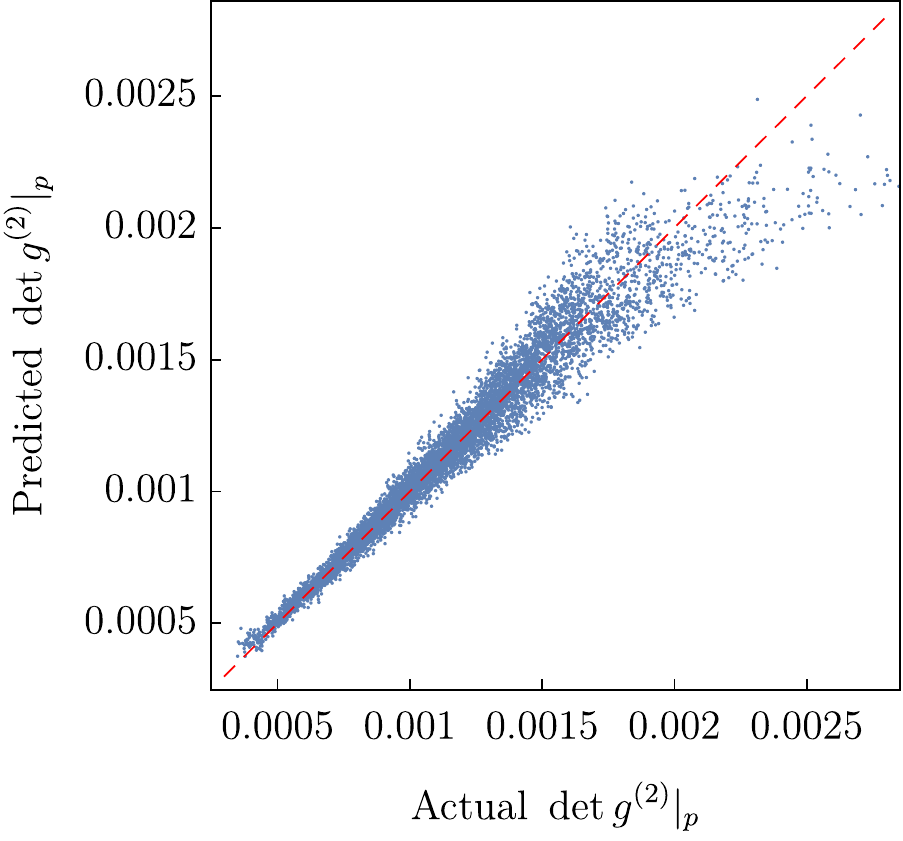}
		\caption{}
			\label{f:detgk=1tok=2a}

	\end{subfigure}%
	\begin{subfigure}{.5\textwidth}
		\centering
		\includegraphics[width=7cm]{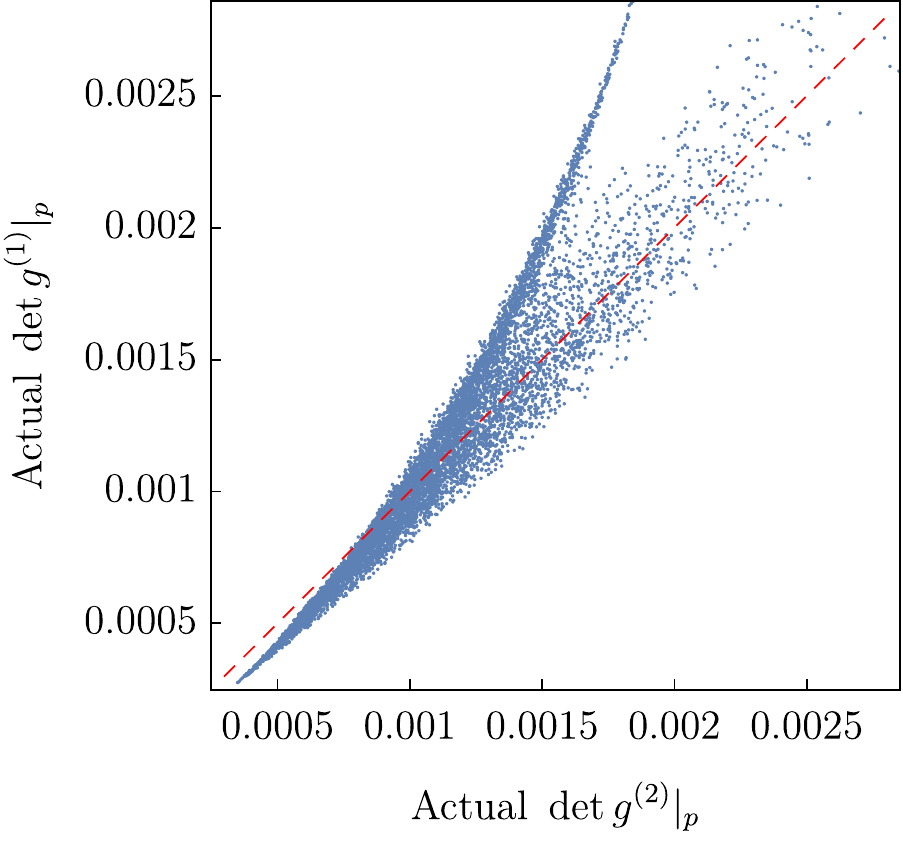}
		\caption{}
			\label{f:detgk=1tok=2b}

	\end{subfigure}
	\caption{Figure (a) shows a scatter plot of the values of $\detgp 2$ versus $\detg 2$. Each point on the plot corresponds to one of the 8,000 random points on $Q$ in the validation set $\cV$. Perfect prediction would correspond to all points lying on the red dashed $y=x$ line. Figure (b) compares the values of $\detg1$ and $\detg2$ for each point -- it is clear to see that the predicted values $\detgp2$ are better than $\detg1$ at approximating the actual values of $\detg2$.}

\end{figure}

Suppose we take a training set $\cT$ of 2,000 random samples from $\cD_{1,2}$. Our validation set $\cV$ will be the remaining 8,000 samples. The ML is trained on $\cT$, the 2,000 samples of points on the quintic with their associated values of $\detg 1$ and $\detg 2$. Once trained, we present it with the remaining 8,000 samples of $\{p , \detg 1|_p \}$ from the validation set $\cV$, and use it to predict the values of $\{\detgp{2}|_p\}$ for those points. We then want to compare the set $\{\detgp{2}|_p\}$ with the known values in  $\{\detg 2|_p\}$ for the sample of 8,000 points. This comparison is shown graphically 
in Figure \ref{f:detgk=1tok=2a}, where we compare the 8,000 values of $\detgp 2$ predicted by the ML versus the actual values of $\detg 2$ computed from the balanced metric. One sees that the predicted values $\detgp 2$ are indeed a good approximation to the actual values of $\detg 2$, with the points clustered around the $y=x$ line without any obvious bias. The best fit curve is
\begin{equation}\label{k=1k=2fit}
y = 0.000098  + 0.92\, x,
\end{equation}
where perfect prediction corresponds to $y=x$. We also compare in Figure \ref{f:detgk=1tok=2b} the values of $\detg 2$ and $\detg 1$, both computed using the balanced metrics -- this is the distribution one would see if the ML were simply using the input value of $\detg 1|_p$ as its predicted value $\detgp 2|_p$. We note that this shows a large deviation away from the $y=x$ perfect-prediction line, indicating that simply taking $\detgp 2|_p=\detg 1|_p$ is worse than the ML. In other words, having seen only 2,000 samples of data, the ML has learned to predict the values of $\detg 2$ for the remaining 8,000 points with impressive accuracy and confidence, all in a matter of seconds.

Our comparison of the actual values versus predicted values, though reassuring, is rather primitive. The linear coefficient in \eqref{k=1k=2fit} indicates only how good the prediction is for \emph{values} of the determinant at $k=2$. What we are really interested in is how close the predicted metric is to the honest Ricci-flat metric (which one would find in the limit $k\to\infty$). A good measure of this is the $\sigma$ error measure, given in \eqref{eq:sigma_int}. Recall that $\sigma$ is determined\footnote{Together with weights, etc., that do not change with $k$ -- for more details, see Appendix \ref{app:donaldson}.} by the values of $\detg k|_p$  summed over the random points of the quintic ($N_{t}=N_g=\text{10,000}$ points in this case). This allows us to compare the $\sigma$ measures computed from $\detg 1$, $\detg 2$ and the predicted values $\detgp 2$, which we will denote by $\sigma_1$, $\sigma_2$ and $\hat\sigma_2$ respectively.

We find 
\begin{equation}
\sigma_1=0.375301,\qquad \sigma_2=0.273948,\qquad\hat{\sigma}_2 = 0.295468.
\end{equation}
The error measure $\hat{\sigma}_2$, computed using the predicted values $\detgp 2$, is significantly smaller than $\sigma_1$ and within 10\% of the actual value of $\sigma_2$. This tells us that our ML provides a much better approximation to the determinant of the Ricci-flat metric than $\detg 1$, and it is relatively close to $\detg 2$ in its accuracy.

Note that since Donaldson's algorithm starts from a K\"ahler potential, the resulting balanced metric is guaranteed to be K\"ahler up to the numerical precision we are working with. One might worry that the predicted values $\detgp{2}$ (or $\hat{g}_{a\bar b}^{(k)}$ if one were predicting the components of the metric) no longer correspond to an \emph{exact} K\"ahler metric. This will indeed be the case since we are predicting the \emph{values} of $\detgp{2}$ and, hence, its ``K\"ahlerness'' is no longer built in. However, given the results of Figure \ref{f:detgk=1tok=2a} and other checks (such as comparing $\vk{}$ calculated using both $\detgp{2}$ and  $\detg{2}$), one can be confident that the underlying predicted metric is still approximately K\"ahler. This also holds true for the other calculations in this paper.

\subsection{Varying the input and output}

Having seen that our ML can learn to predict the values of $\detgp 2$ from a small sample of $\detg2$ data, it is natural to ask whether it can repeat this for higher values of $k$. That is, can the ML learn to predict $\detg k$, where $k>2$, from $\detg1$? 

\begin{figure}[!h!t!b]
	\centerline{
		\hspace{-3em}\includegraphics[trim=0mm 0mm 0mm 0mm, clip, width=10cm]{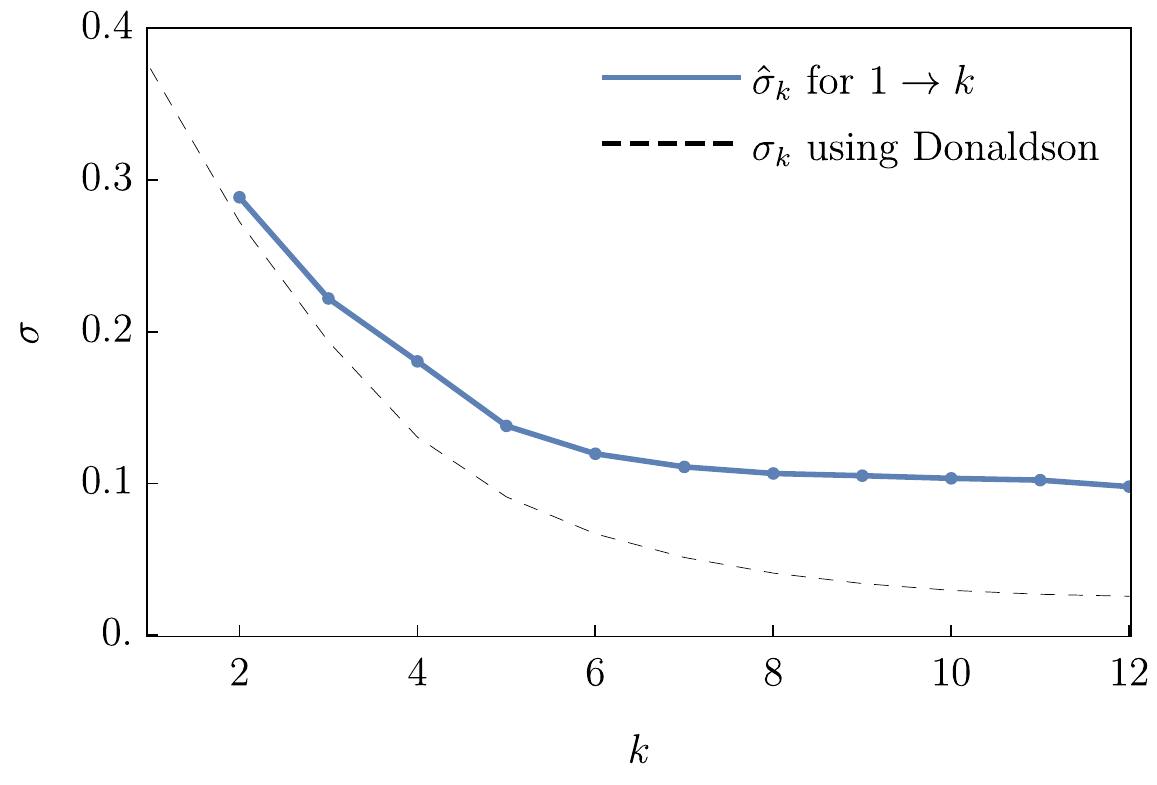}
	}
	\caption{Comparison of $\hat\sigma_k$ -- computed using the values of $\detgp k$ predicted from $\detg 1$ -- with $\sigma_{k}$ computed using Donaldson's algorithm. 
		\label{fig:1_to_k}}
\end{figure}

For this experiment we fix $N_g=\text{20,000}$ random points on the quintic and split them into training and validation sets, each of size 10,000. In the notation of \eqref{detgData}, for each $2<k<12$ we take $\mathcal{D}_{1,k}$ to be 20,000 samples of data and split it into a training set $\cal{T}$ and validation set $\cal{V}$, each of size 10,000. For each value of $k$, up to $k=12$, we train an ML on $\{p , \detg 1|_p \}\to\{\detg k|_p \}$ in $\cal{T}$. Using the ML, we predict the values of $\detgp k|_p$ at each point for the 10,000 validation samples in $\cal{V}$ and compute the resulting $\sigma$ error measure. We plot the predicted values $\hat\sigma_k$ in Figure \ref{fig:1_to_k}. We see that when the ML is trained on higher k data (as $k$ increases, the balanced metrics are closer to Ricci-flat), its predictions for $\detgp k$ result in smaller error measures. 
We note, however, that the improvement plateaus around $k=7$, suggesting that the information contained in $\{p,\detg 1|_p\}$ is not sufficient to predict the values of $\detg{k\gtrapprox7}$ with greater accuracy.

Having seen that one can use ML to predict the determinant at higher degrees by training it on a small set of training data, consisting of both $k=1$ and the higher $k$ values, we now explore how the accuracy of our routine changes when we increase the degree used to compute the input data, replacing $\detg1|_p$ with $\detg l|_p$ for $l>1$. 
For example, consider training an ML to predict the values of $\detg 4$. We might try to predict $\detg 4$ from $\detg 2$ instead of $\detg 1$. 

\begin{figure}[!h!t!b]
	\centerline{
		\hspace{-5em}\includegraphics[trim=0mm 0mm 0mm 0mm, clip, width=10cm]{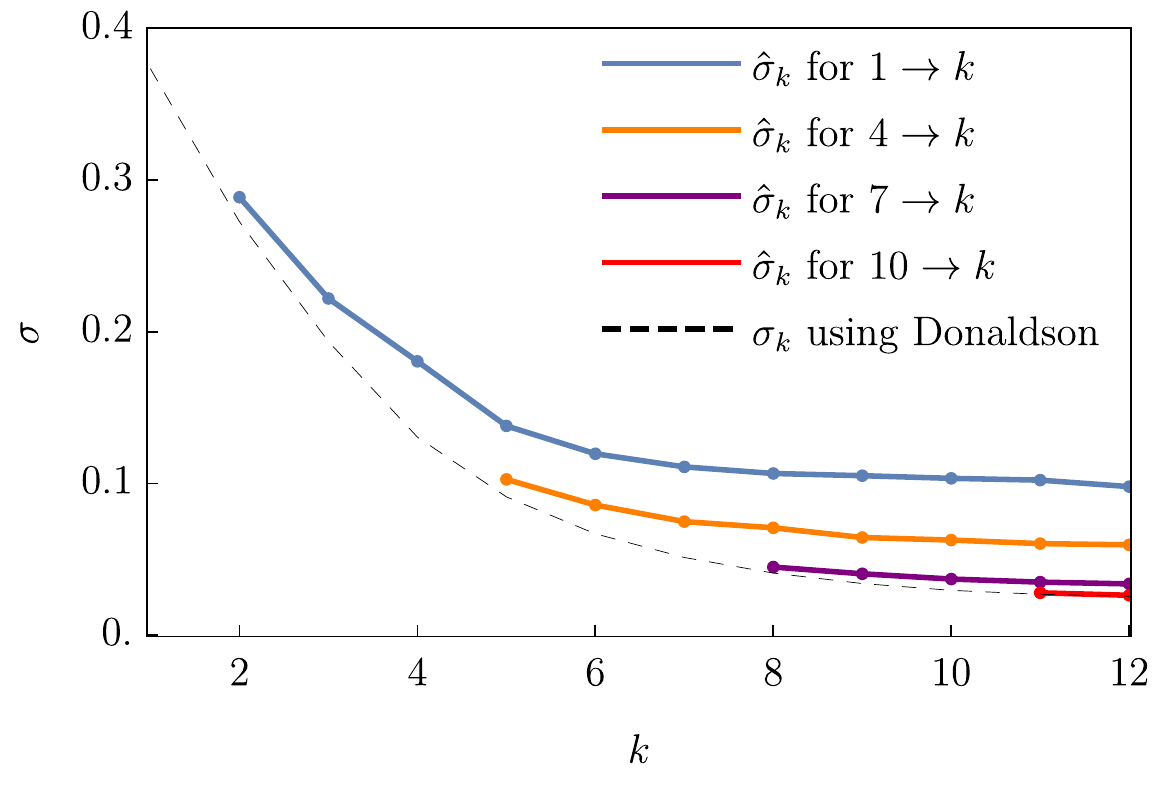}
	}
	\caption{Comparison of values of $\hat\sigma_k$ where we vary the input determinant. For example, the blue line denotes the $\hat\sigma$ values calculated from predicting the value of $\detg k$ from $\detg 1$; the orange line denotes the $\hat\sigma$ values calculated from predicting the value of $\detg k$ from $\detg 4$; and so on. The dashed black line indicates the values of $\sigma_k$ computed using the balanced metric obtained from Donaldson's algorithm.
		\label{fig:sigma_increase_degree}}
\end{figure}

Again, we fix $N_g=\text{20,000}$ random points on the quintic and split them into training and validation sets, each of size 10,000. We then train an ML on $\{p , \detg l|_p \}\to\{\detg k|_p \}$ in $\cal{T}$, one ML for each pair of $l$ and $k$, with $l=4,7,10$ and $k=l+1,\ldots,12$. Using the ML, we predict the values of $\detgp k|_p$ at each point for the 10,000 validation samples in $\cal{V}$ and compute the resulting $\sigma$ error measure. In Figure \ref{fig:sigma_increase_degree} we plot the predicted values of $\hat\sigma_k$ as we vary the degree for the input determinant. For example, we see that training the ML on $\mathcal{D}_{1,12}$ (using $\detg 1$ as input and $\detg{12}$ as output) leads to a larger $\sigma$ measure than $\mathcal{D}_{7,12}$ (using $\detg 7 \to \detg{12}$ data). As might be expected, the ML's predictions are better 
(where ``better'' is measured by how close the predicted $\hat\sigma_k$ is to $\sigma_k$, the error measure computed using Donaldson's balanced metric) when the degree of the input determinant $l$ is larger.



\subsection{Comments}

As we have seen in this section, an ML is able to learn the determinant of the balanced metric for a total labelled data set $\cal{D}$ having seen only a small amount of the data given in the training set $\cal{T}$. This provides an important check on our approach. However, as discussed above, the method described in subsections 3.3 and 3.4 still requires the use of Donaldson's algorithm to compute the values of $g^{(k)}$, albeit for a smaller set of training points. Hence, it remains necessary to calculate $\hmat$ at the higher value of $k$ -- a very time-consuming procedure that becomes factorially slower as the value of $k$ increases. 
In the next few sections, we will discuss how to modify our machine-learning algorithm so as to remove the need for the sample data $\{\detg k|_p\}$ at the higher value of $k$.
We will then compare this new machine-learning algorithm with the known results from the balanced metric. In practice, when one is trying to extend calculations to degrees $k$ that are too large for Donaldson's algorithm to finish in a reasonable time, one will not have the balanced metric to compare with. Thus it is important that we are confident that our supervised-learning model is trustworthy.

Note that the results of this section are of interest on their own -- Calabi--Yau metrics (and the balanced metrics that approximate them) are algebraically complicated, so for those not familiar with machine learning it might be surprising that we can achieve such accuracy with so small an amount of training data. Again and again, machine learning has proved able to learn complicated algorithms or infer behaviour from data without any known unified mathematical description. The exact way it does this is often obscure -- we are not able to offer any insight into why our data is so amenable to ML.

While Donaldson's algorithm is factorial in complexity with respect to $k$ (the size of the monomial basis, the size of the matrix $\hmat$ and the number of points $N_p$ all increase factorially), the machine-learning approach, which focuses only on the final result of $\detg k$ as a distribution over the random points, does not grow in complexity. This makes machine learning extremely attractive from a speed point of view. One could well imagine packaging a trained ML to allow researchers to do their own calculations using Calabi--Yau metrics without having to go through the entire process of calculating the balanced metric, and so on.

As we have mentioned many times, the nature of supervised learning means that the ML has to be trained on a sample of values of $\detg k$ in $\cal{T}$ computed at the higher value of $k$. To obtain these, one could follow Donaldson's algorithm for computing the balanced metric and then compute at least some values of $\detg k$, as we did above. Ideally, however, one would like to avoid this calculation entirely, side-stepping the need to compute the balanced metric for the higher value of $k$. In the following section, we present a simple extrapolation approach that does just this. In the section after that, we combine this extrapolation with our machine-learning model to quickly obtain accurate predictions for the determinant without having to compute the balanced metric at the higher value of $k$.

\section{Extrapolating out to higher \texorpdfstring{$k$}{k}}\label{sec:curve_extrap}

In the previous section, we saw that supervised learning provides a quick and accurate way to obtain properties of the metric (such as the value of the determinant) for all points using only the data of a small number of training points.\footnote{ 
	The idea of using a small sample of difficult-to-compute quantities to ``seed'' an ML was used in \cite{Bull:2019cij} to predict Hodge numbers of more complicated Calabi--Yau manifolds from simpler ones.}
Unfortunately, as we noted above, one still needs to compute the values of $\detg k$ at the higher $k$ value for the small number of training points -- this is simply the nature of supervised learning.

We now discuss how one can obtain a similar result without needing to calculate $\detg k$ for the higher values of $k$. We do this using a simple extrapolation based on regression and curve fitting. We will see that given the values of $\detg k$ for a small range of $k$ values, one can accurately extrapolate to higher values of $k$. On its own, this provides a way to obtain more accurate numerical values of $\detg k$, side-stepping Donaldson's algorithm and the need to find the balanced metric. Unfortunately, curve fitting for a large number of points, say $N_g=\text{500,000}$, is still very time consuming. To mitigate this problem, in the next section, we will combine curve fitting with machine learning: curve fitting will be used to obtain the training data on a relatively small number of points, and then the previously discussed supervised-learning routine can be used to predict the values of $\detg k$ for all 500,000 points. Together, this gives a substantial speed up compared with following Donaldson's algorithm for larger $k$ values.

\begin{figure}[!h!t!b]
	\centerline{
		\hspace{-5em}\includegraphics[trim=0mm 0mm 0mm 0mm, clip, width=10cm]{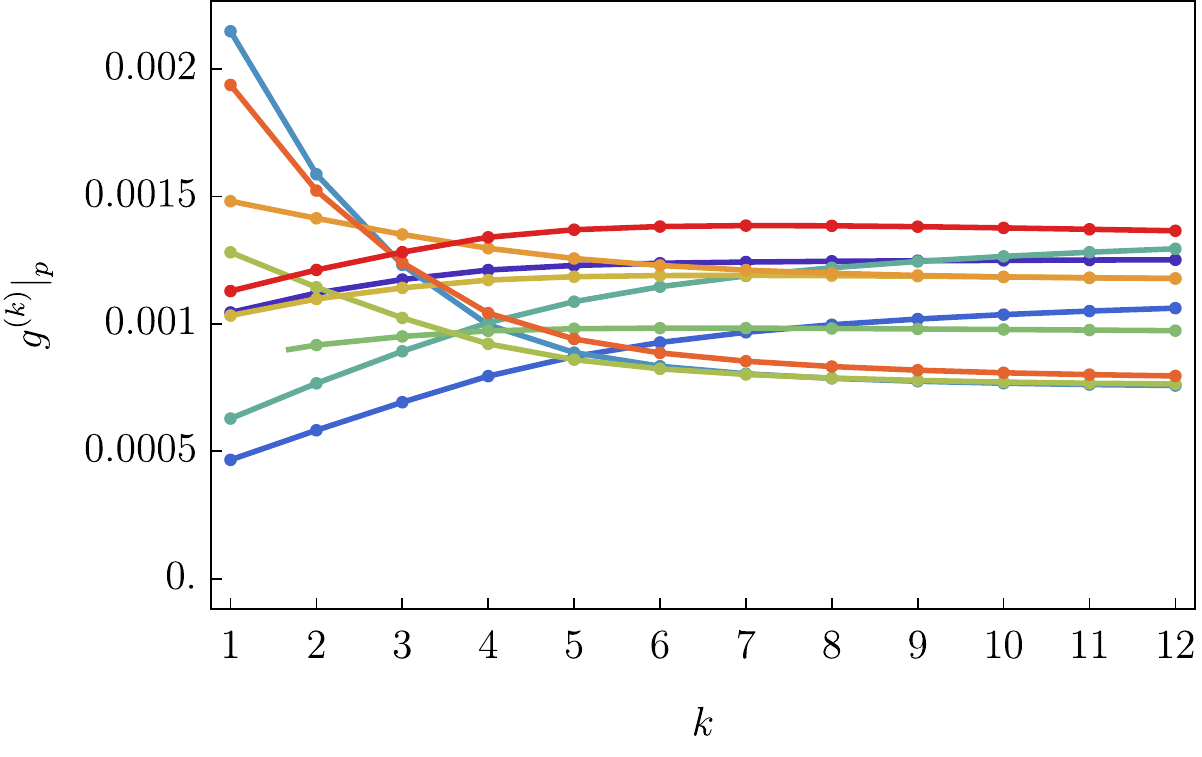}
	}
	\caption{A plot of the value of $\detg k$ computed, using Donaldson's algorithm, from the balanced metric as a function of $k$ for ten randomly chosen points on the quintic. We see the generic behaviour is that of a (rising or falling) decaying exponential -- the values tending to constants for large $k$.
		\label{fig:det_as_fn_k}}
\end{figure}

As before, we will focus on a scalar quantity, namely the determinant of the metric. Donaldson's algorithm produces the balanced metric for each chosen degree $k$. Thus, for every point $p$ on the quintic, one can compute, using Donaldson's algorithm, the determinant $\detg k$ for each degree $k$. In this way, we have a list of values of $\detg 1$, $\detg 2$, and so on, for each point $p$ on the manifold. The idea is to examine the behaviour of $\detg k |_p$ for each individual point as $k$ varies. In Figure \ref{fig:det_as_fn_k}, we show how the value of $\detg k$ changes with $k$ for ten randomly selected points. We see that the behaviour of the determinant for each point can be well approximated by a decaying exponential plus a constant term; that is
\begin{equation}\label{eq:curve_ansatz}
\detg k |_p = a_p - b_p\ee^{-c_p k},
\end{equation}
where $a_p$, $b_p$ and $c_p$ are fixed parameters that depend on the choice of point $p$.

The form of this equation is not entirely surprising and makes intuitive sense. As $k\to\infty$, the balanced metric that Donaldson's algorithm produces gets closer and closer to the honest Ricci-flat metric. Similarly, the value of $\detg k$ evaluated at a point on the manifold must also tend to the value corresponding to that of the Ricci-flat metric $g$. The precise way that $\detg k$ tends to its final value is certain to be complicated, but it is clear that, other than at particularly singular points, it should approach its asymptotic value in a relatively smooth manner. Moreover, the rate at which it tends to its final value should be such that for each extra degree in $k$, there is a diminishing gain in the accuracy of $\detg k$ (as measured by evaluating $\sigma_k$). Together, these suggest a decaying exponential with a constant shift would be a reasonable description of how $\detg k$ changes with increasing $k$.

\begin{figure}[!h!t!b]
	\centerline{
				\hspace{-3.5em}\includegraphics[trim=0mm 0mm 0mm 0mm, clip, width=10cm]{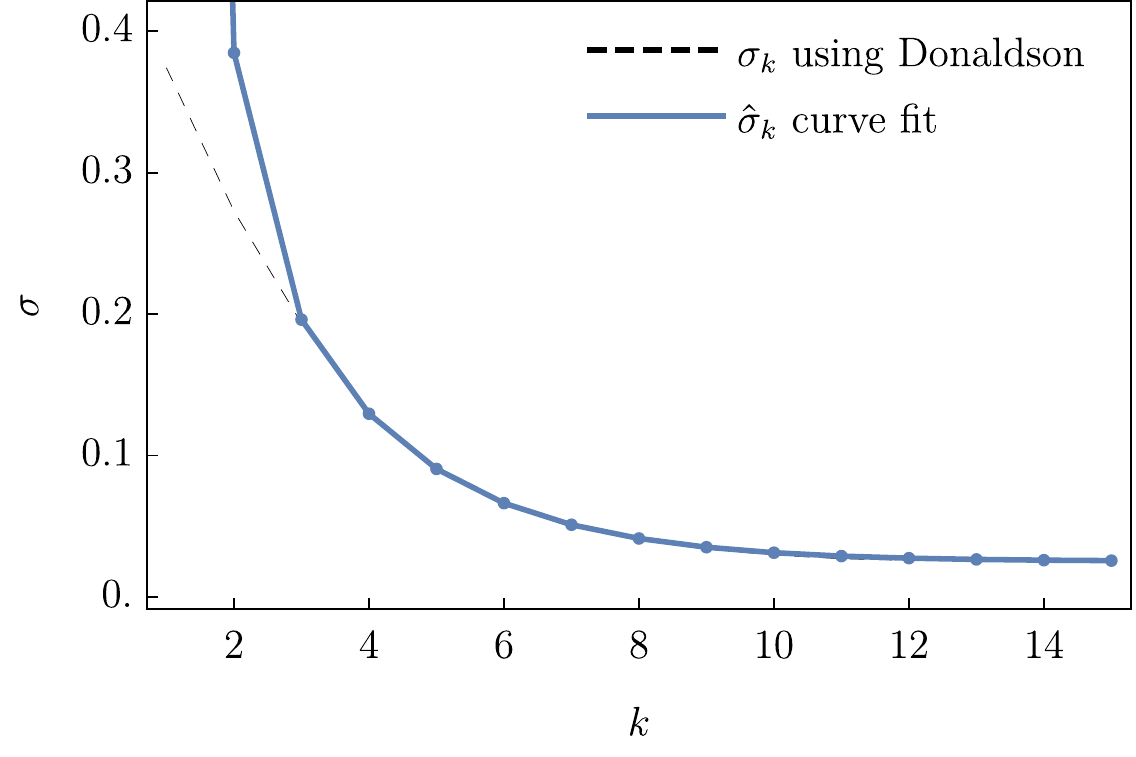}
	}
	\caption{A comparison of the $\sigma$ measure computed for $N_t=\text{10,000}$ points using: a) the known values of $\detg k$ computed using the balanced metric; b) the predicted values of $\detg k$ computed using curve fitting on the $k=4,\ldots,7$ values. \label{fig:sigma_curve_fit}}
\end{figure}

One simply fits an equation of the form \eqref{eq:curve_ansatz} to the values of $\detg k|_p$ for each point $p$ on the manifold.\footnote{In practice, one can use Mathematica's \texttt{Fit[]} function. This finds the values of $a_p$, $b_p$ and $c_p$ that minimise the sum of the squared differences between the actual values of $\detg k |_p$ and the values given by $a_p - b_p \ee^{-c_p k}$.} To check whether this proves to be useful, we compute the $\sigma$ measure. Using only the values of $\detg k$ for $k=4,5,6,7$ to fit the curves, we can predict the values of $\detgp k$ for higher $k$. One can then use these predicted values to compute $\hat\sigma_k$ and compare this with that calculated using Donaldson's algorithm. For this comparison, it is sufficient to take $N_{t}=\text{10,000}$. As we see in Figure \ref{fig:sigma_curve_fit}, the $\sigma$ measures are remarkably close, suggesting both that our ansatz \eqref{eq:curve_ansatz} for the behaviour of $\detg k$ is reasonable and that $\detg k$ behaves relatively smoothly as $k$ increases. Note that the input for this calculation is the points and the values of $\detg k|_p$ for $k=4,\ldots,7$. No values for higher $k$, $\detg{12}$ for example, are used. It is encouraging to see the predicted value $\hat\sigma_k$  matching that computed using the balanced metric all the way up to $k=12$ (as far as we have pushed Donaldson's algorithm) and even slightly beyond. The ``best'' prediction of $\detgp k |_p$ is given by taking $k\to\infty$ in \eqref{eq:curve_ansatz}, leaving only the constants $a_p$ which give the predicted asymptotic value of $\detgp k |_p$ at a given point $p$. We denote the corresponding value of the error measure by $\hat\sigma_\infty$.

One might wonder why we have picked the range $k=4,\ldots,7$ as an input for the curve fitting. As we will see in the next section, this range results in predictions that are equivalent in accuracy to the $k=12$ balanced metric computed using Donaldson's algorithm. This allows us to directly compare the calculation times that one needs to achieve the same accuracy, that is the same $\sigma$ measures. In practice, one will not know in advance what kind of accuracy one will achieve with a given range of input $k$ values. Instead, the range might be chosen by deciding how much time one is willing to spend calculating the input data. For example, one might compute $k=4,\ldots,8$ instead, which will take longer to calculate but will lead to better curve fitting and a lower predicted $\hat\sigma$ error measure. For the remainder of this paper, we stick with $k=4,\ldots,7$ as the input data for curve fitting.

Recall that we are actually interested in predicting the determinant for $N_g=500{,}000$ points rather than the smaller sample of $10{,}000$ we have considered here. Unfortunately, fitting a curve for each of $500{,}000$ points is both time and resource hungry -- curve fitting in this manner does not easily scale. Machine learning, however, is well suited to problems with large data sets. Our plan is to use this ``curve extrapolated'' data to provide the small amount of ``seed'' output data for the training set $\cal{T}$ for our previous supervised-learning model. The idea is that we use Donaldson's algorithm to compute the balanced metric, and thus $\detg k$, for just $k=4,5,6,7$. We can then extrapolate to find the values of $\detgp k|_p$ out to $k=12$ for a small number of points, say 10,000, and use this data as an input to the training set $\cal{T}$ of our previous supervised-learning model. We can then train this model to estimate the $k=12$ values for the rest of the $N_g=\text{500,000}$ points. If on wants to obtain the ``best'' predictions of $\detgp k|_p$, one takes $k\to\infty$ in \eqref{eq:curve_ansatz}, resulting in a predicted error measure $\hat\sigma_\infty$ that can be compared with $\sigma_k$ computed using Donaldson's balanced metric. As we will see, this provides a quick way to compute $\detg k$ without sacrificing much in the way of accuracy.

\section{Supervised learning and extrapolation: results}\label{sec:learn_extrap}

In the previous two sections, we have explored how a particular property of an approximation to a Calabi--Yau metric, namely the determinant, can be captured by machine-learning or simple curve fitting. Let us remind ourselves of one of the goals stated in the introduction. If we are to use string theory to make contact with particle physics, we must be able to compute masses, couplings, and so on, from first principles. As a start, this will involve computing correctly normalized cubic couplings. To do this, we would like to have a robust and relatively quick numerical scheme for computing quantities associated with Calabi--Yau metrics. Practically speaking, this means being able to compute the metric, its determinant, zero modes of the Laplacian, and so on, to high accuracy without needing a supercomputer.

We have already seen that, given a small sample of training data $\cal{T}$, we can train an ML to accurately and quickly predict values for the determinant of the metric for the remaining points in the validation set $\cal{V}$. This is wonderful in principle but does not help us much in practice -- we still have to ``seed'' the training data with some of the higher-accuracy (higher $k$) data. This requires using Donaldson's algorithm to compute this higher $k$ training data, which is computationally expensive and extremely time consuming for large value of $k$. In the previous section, we saw that one can actually extrapolate from lower $k$ out to higher $k$ using simple curve fitting. Unfortunately, this kind of fitting is also very slow in practice and not suited to computations with $N_g=500{,}000$ points, as are needed when computing Laplacian eigenmodes, for example.

The idea of this section is to combine both of these approaches to obtain accurate predictions for the determinant that are much faster than each of the above individual methods and, hence, useful in practice. Using the $\detg k$ data computed for low values of $k$, we will use curve fitting to extrapolate out to larger values of $k$. We have to do this for only a small number of points, say 10,000, since the ML needs only a small amount of data to be trained (as we saw in Section 3). We can then use the extrapolated values of $\detgp k$ as the outputs in the training set. Using supervised learning, as in Section 3, we train an appropriate ML to quickly predict the values of $\detg k$ at the higher value of $k$ for the remainder of the $N_g=\text{500,000}$ points.

\begin{figure}[!h!t!b]
	\centerline{
		\hspace{-3.5em}\includegraphics[trim=0mm 0mm 0mm 0mm, clip, width=10cm]{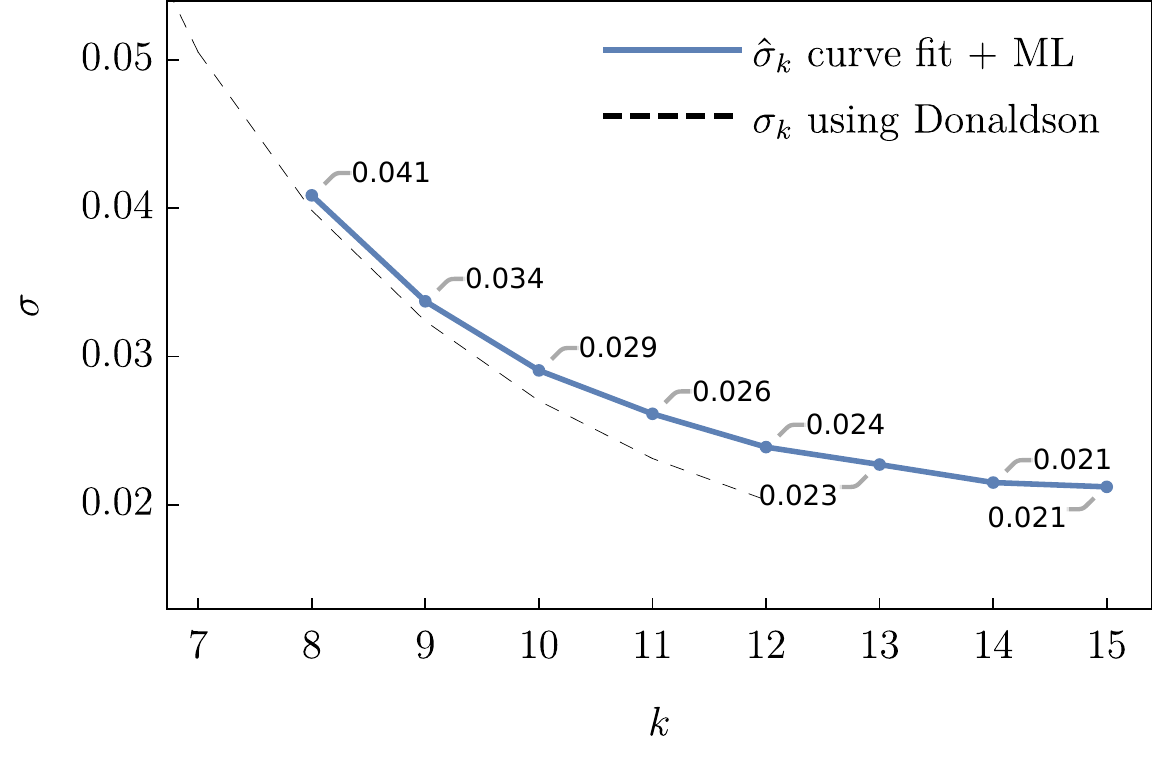}
	}
	\caption{A comparison of the $\sigma$ measure computed for $N_t=N_g=\text{500,000}$ points using: a) the predicted values $\detgp {k}$ using an ML trained on the values of $\detg k$ determined from curve fitting using $k=4,5,6,7$ -- the solid blue curve; b) the values of $\detg k$ computed using the balanced metric from Donaldson's algorithm -- the dashed black curve. \label{fig:sigma_curve_ML}}
\end{figure}

Let us lay out explicitly the steps we will follow:
\begin{enumerate}
	\item We fix $N_g=500{,}000$ points on the quintic for which we would like to compute $\detg k$ to high accuracy.
	\item We use Donaldson's algorithm to compute the balanced metric $\hmat$ at $k=4$. Note that, from \eqref{eq:point_bound} and \eqref{Nk}, 
we need only $N_p=99{,}000< 500{,}000$ points to evaluate the $T$-operator as we do not need all 500,000 for convergence to the balanced metric. 
	\item Using this $\hmat$, we compute the values of $\detg 4$ for \emph{all} $N_g=\text{500,000}$ points.
	\item We repeat the previous two steps for $k=5,6,7$ which, from  \eqref{eq:point_bound} and \eqref{Nk}, require $N_p=206{,}250$, $N_p=470{,}250$ and $N_p=1,042{,}250$  points respectively to evaluate the $T$-operator. Using the resulting $\hmat$ matrices, we compute the values of the determinant for $k=5,6,7$ for all 500,000 points. After this step, we have the values of $\detg k$ for $k=4,5,6,7$ at all $N_g=\text{500,000}$ points.
	\item Select a subset of $10{,}000\subset500{,}000$ points along with their values of $\detg k$ for $k=4,5,6,7$. Using the curve fitting approach discussed in the previous section, we predict the values of $\detg k$ for each point up to a larger value of $k$. This gives us 10,000 extrapolated values of $\detgp{15}$ that we can use as an input data to train an ML.
	\item We train an ML (using the approach outlined in Section 3) using 10,000 samples of the form
	\begin{equation}
	\{p,\detg{4}|_p,\detg{5}|_p,\detg{6}|_p,\detg{7}|_p\}\to\{\detgp{k}|_p\},
	\end{equation}
	where $p$ is the affine coordinate of a point, $\detg{4,5,6,7}$ are the determinants computed from the balanced metric and $\detgp{k}$ are the values given by the curve fitting. Since we already have the values of $\detg k$ for $k=4,5,6,7$, we may as well include them when training the ML. This input data forms our total training set $\cal{T}$.
	\item We now have an ML that can be used to quickly predict the values of $\detgp{k}$ for the $490{,}000=500{,}000-10{,}000$ remaining validation samples $\cal{V}$. We already have the points and values of $\detg{4,5,6,7}$ for the remaining samples, from which the trained ML is able to predict $\detgp{k}$. 
	\item Using the predicted values $\detgp{k}$ for all $N_g=\text{500,000}$ points, we can compute $\hat{\sigma}_{k}$ to check the accuracy of the predictions.
\end{enumerate}
Following these steps, we have computed the $\hat{\sigma}_{k}$ values from the predicted values of $\detgp {k}$ for $k=8\ldots,15$, which we show in Figure \ref{fig:sigma_curve_ML}. We also plot the values of $\sigma_{k}$ computed using the balanced metric itself (as in Section 2).

Taking $k\to\infty$ in $\detgp k$ to obtain the ``best'' possible prediction, combining curve fitting and machine learning then gives a predicted error measure $\hat\sigma_\infty$ equal to that of directly computing the balanced metric at $k=12$; one finds
\begin{equation}
\hat\sigma_\infty = \sigma_{12}
\end{equation}
to approximately 2\%. This means we should compare our combined curve fitting and machine learning approach with calculating the balanced metric at $k=12$. Remember that while the latter forces us to follow Donaldson's algorithm for $k=12$, the combined curve-fitting and machine-learning method only requires us to use Donaldson's algorithm for $k=4,5,6,7$. Note also that the value of $\hat\sigma_{\infty}$ is much smaller than $\sigma_7$ with more than a factor of two improvement (recall that $k=7$ is the most accurate balanced metric that one must compute for the curve fitting). As a sanity check, we also computed the volumes $\vk{}$ defined by $\detg{12}$ and $\detgp{\infty}$ via \eqref{eq:vols} and found agreement to better than 0.1\%.

Most importantly, since the time that Donaldson's algorithm takes scales factorially with $k$, it turns out that our new combined method is \emph{much} quicker -- let us put some numbers on this. As we saw in Figure \ref{fig:h_time} in Section 2, following Donaldson's algorithm with $k=12$, one can find the balanced $\hmat$-matrix in approximately 182,000 seconds. Given $\hmat$, one can then calculate the values of $\detg{12}$ for the $N_g=\text{500,000}$ points of interest in another 6,000 seconds, giving a total runtime of 188,000 seconds (or 52 hours). If, instead, we combine curve fitting and machine learning we have to sum: 1) 900 seconds to find $\hmat$ for $k=4,5,6,7$, with 2) 1,400 seconds to calculate the values of $\detg k$ for $k=4,5,6,7$ for all 500,000 points, with 3) 130 seconds to curve fit $\detg k|_p$ and extrapolate out to $k\to\infty$ for 10,000 points; and, finally with 4) 70 seconds to train an ML using the extrapolated data and predict $\detgp{\infty}$ for the remaining 490,000 points. This gives a total time of approximately 2,500 seconds (42 minutes). That is, for $N_g=\text{500,000}$ random points on the quintic
\begin{equation}
\begin{split}
\text{Time to find }\detg{12}\text{ via Donaldson: }&188{,}000\text{ seconds},\\
\text{Time to find }\detgp{\infty}\text{ via curve fitting + ML: }&2{,}500\text{ seconds}.\\
\end{split}
\end{equation}
Comparing the two times, we see that utilizing curve fitting and machine learning leads to a speed-up by a factor of 75, almost two orders of magnitude. 

It is interesting to ask: rather than using both curve fitting and machine learning, might one simply use curve fitting alone. That is, one could simply predict the values of $\hat{g}^{(\infty)}$ for all 500,000 points. Unfortunately, as we mentioned in Section \ref{sec:curve_extrap}, this is rather slow and much slower than using machine learning. If we had used curve fitting alone, the timings would be: 1) 900 seconds to find $\hmat$ for $k=4,5,6,7$, with 2) 1,400 seconds to calculate the values of $\detg k$ for $k=4,5,6,7$ for all 500,000 points, with 3) 6,500 seconds to curve fit $\detg k|_p$ and extrapolate out to $k\to\infty$ for 500,000 points. This would give a total time of approximately 8,800 seconds (146 minutes). This is still a factor of 21 faster than using Donaldson's algorithm for $k=12$, but is 3.5 times slower than combining curve fitting for a small sample of points and using machine learning to predict the rest. We conclude that the algorithm introduced in this paper which combines both machine learning with curve fitting a small number of data points, is the least time consuming and most efficient approach to computing $\detg k$.

As we have mentioned, in practice one would like to find the values of $\gk$ rather than $\detg k$. The timings for Donaldson's algorithm would remain unchanged (since $\gk$ is computed in order to find $\detg k$). For our combined approach, the curve fitting and machine learning contributions to the timing would increase by a factor of nine or so (since $\gk$ has nine independent components, each of which must be predicted). In this case, our method would give a speed-up by a factor of 50 or so. In the following section we show that the components of the complete Calabi--Yau metric can indeed be rapidly computed to high accuracy using our combined algorithm.

\section{Predicting the metric}\label{sec:full_metric}

We have now seen that, using a combination of curve fitting and machine learning, it is possible to predict the values of the determinant for 500,000 points on the quintic in much less time than Donaldson's algorithm alone whilst achieving similar accuracy. However, it is not the determinant we are really interested in; we really want the metric itself since this enters numerical calculations of gauge connections and harmonic modes. As we have emphasized throughout the paper, everything we have done for the determinant applies equally well to the components of the metric itself. As we noted in subsection \ref{sec:learn_det}, for each patch on the quintic the hermitian metric $g_{a\bar b}(x,\bar x)$ can be thought of as a collection of nine independent real functions. Each of these functions can be predicted using exactly the same approach we adopted for the determinant.

\begin{figure}[!h!t!b]
	\centerline{
		\hspace{-5em}\includegraphics[trim=0mm 0mm 0mm 0mm, clip, width=10cm]{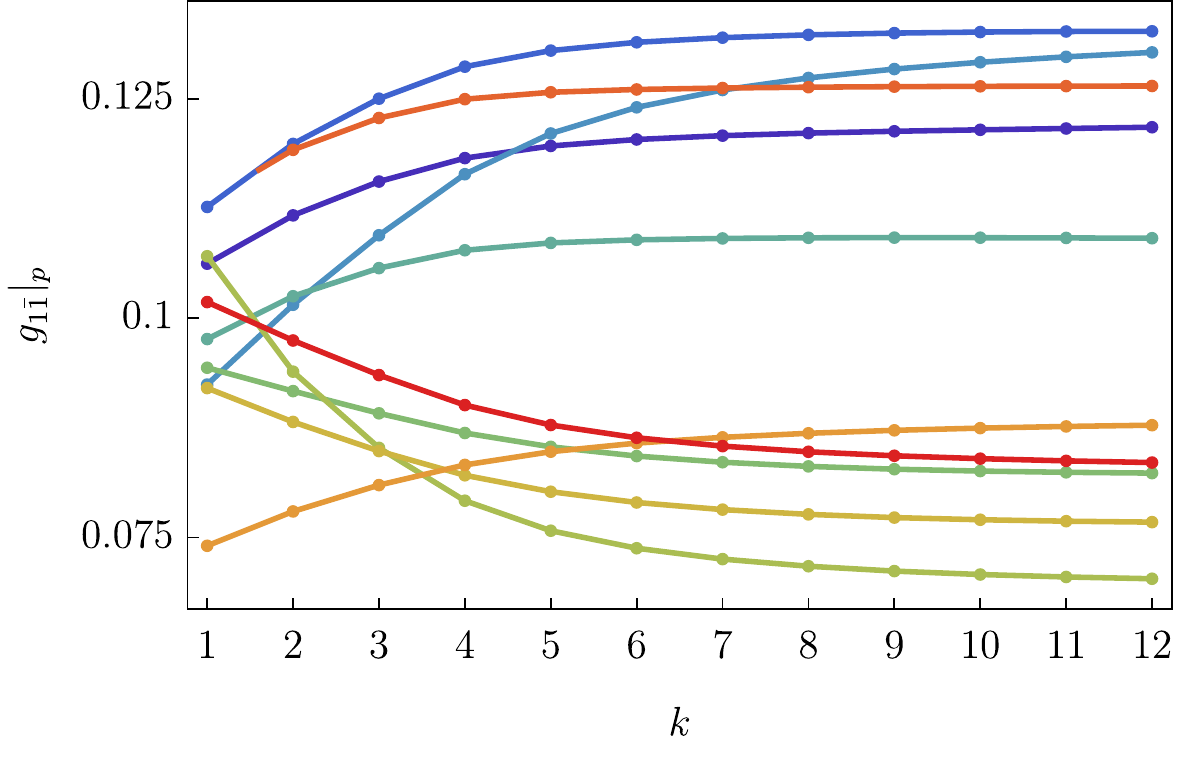}
	}
	\caption{A plot of the value of $g_{1\bar1}^{(k)}$ computed, using Donaldson's algorithm, from the balanced metric as a function of $k$ for ten randomly chosen points on the quintic. We see the generic behaviour is that of a (rising or falling) decaying exponential -- the values tending to constants for large $k$.
		\label{fig:g_11_as_fn_k}}
\end{figure}

For the determinant, our ability to extrapolate out to larger values of $k$ using curve fitting relied on $\detg k|_p$ behaving as in equation \eqref{eq:curve_ansatz}. Thus, to be certain that the same extrapolation will work for the components of the metric, we should check how $\gk |_p$ behaves as $k$ varies. Let us focus on $g_{1\bar1}^{(k)}$ as it is real -- everything we say applies to the real and imaginary parts of the remaining components of $\gk$. In Figure \ref{fig:g_11_as_fn_k} we show how the value of $g_{1\bar1}^{(k)}$, computed using Donaldson's algorithm, changes with $k$ for ten randomly selected points on the quintic. As before with the determinant, we see that the behaviour can be well approximated by a decaying exponential plus a constant term; that is
\begin{equation}
g_{1\bar1}^{(k)}|_p = a^{1\bar1}_p-b^{1\bar1}_p \ee^{-c_p^{1\bar1}k},
\end{equation}
where $a^{1\bar1}_p$, $b^{1\bar1}_p$ and $c^{1\bar1}_p$ are fixed parameters that depend on the choice of point $p$.

\begin{figure}[!h!t!b]
	\centerline{
		\hspace{-5em}\includegraphics[trim=0mm 0mm 0mm 0mm, clip, width=7cm]{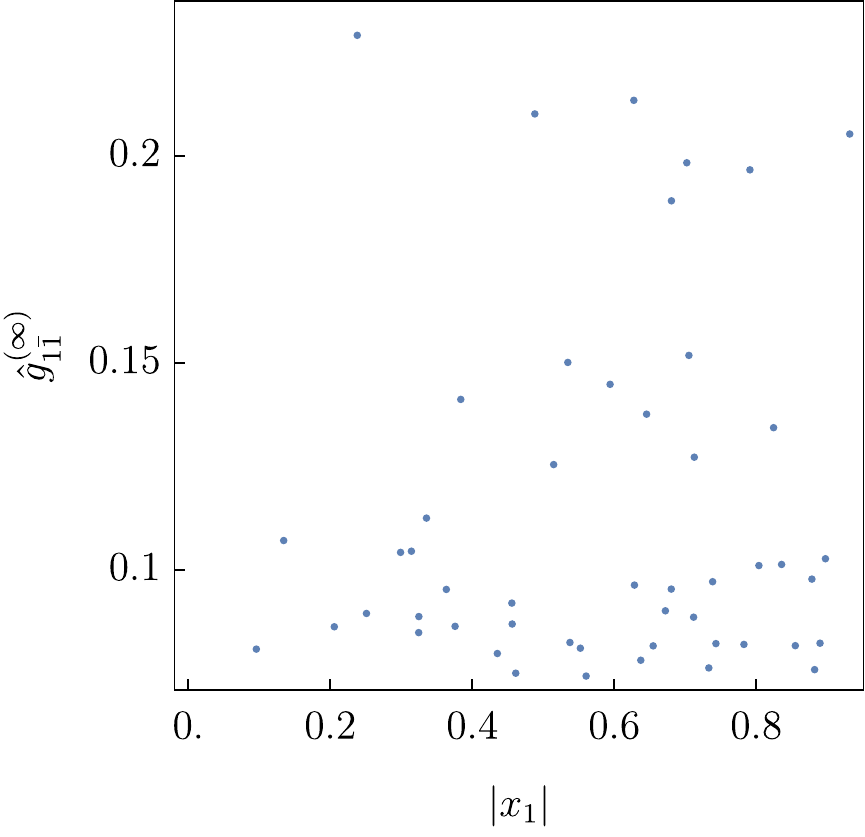}
	}
	\caption{A plot of the predicted value of $\hat{g}_{1\bar1}^{(\infty)}$ as a function of $|x_1|$ (one of the good coordinates on $Q$) for fifty randomly chosen points. As one would expect, there is no structure to the distribution of points.
		\label{fig:g11_scatter}}
\end{figure}

With this result in hand, one can proceed exactly as in the previous section with $\detg k$ and $\detgp k$ replaced with $g_{a\bar b}^{(k)}$ and $\hat{g}_{a\bar b}^{(k)}$. Since one has to compute $h^{\alpha\bar\beta}$ and $\gk|_p$ to find $\detg k|_p$, the timings of the previous section related to Donaldson's algorithm are unchanged -- it again takes 900 seconds to find $h^{\alpha\bar\beta}$ for $k=4,5,6,7$ and another 1,400 seconds to calculate the values of $\gk|_p$ for all $N_g=500{,}000$ points. Since there are nine real degrees of freedom in $\gk$ (since it is hermitian), the time to curve fit (using only the $k=4,5,6,7$ data), train an ML and predict $\hat{g}^{(\infty)}$ for all 500,000 points is simply nine times greater than it was for the determinant; that is, it takes $9\times(70+130)=1{,}800$ seconds. This gives a total time approximately 50 times smaller than using Donaldson's algorithm alone at $k=12$. At the end of this calculation, one has a predicted value $\hat{g}_{a\bar b}^{(\infty)}$ for the metric at each of $N_g=500{,}000$ points on the quintic, given explicitly as a set of 500,000 $3\times3$ numerical hermitian matrices. As an example of this, we plot the value of $\hat{g}_{1\bar1}^{(\infty)}$ as a function of $x_1$ in Figure \ref{fig:g11_scatter} for fifty randomly chosen points. Taking the determinant of each of these numerical matrices, one can then calculate the corresponding $\hat\sigma_\infty$ error measure, finding it is again comparable to the $k=12$ balanced metric. We conclude that the combined curve fitting plus machine learning algorithm introduced, and applied to the determinant, in Section 5, is equally applicable to computing the full Ricci-flat Calabi-Yau metric and does so more than an order of magnitude faster than using Donaldson's algorithm alone.

\section{Discussion}

In this paper, we have applied machine learning to the problem of finding Calabi--Yau metrics. We reviewed how Donaldson's algorithm provides a numerical approximation to the Ricci-flat K\"ahler metric on a Calabi--Yau manifold and pointed out the computational problems associated with pushing the algorithm to higher accuracy. In the hope of speeding up these calculations, we outlined how machine learning, and supervised learning in particular, might be used to predict higher accuracy metrics. To avoid having to input some of the higher accuracy data using Donaldson's algorithm, we suggested a straightforward curve fitting routine that then provides training data for an ML to learn from. Focussing on the determinant of the metric as an example, we showed that one can combine curve fitting and supervised learning to obtain high-accuracy approximations to the Ricci-flat metric starting from lower-accuracy data. Importantly, this approach leads to a great improvement in the speed of calculations, giving a speed-up by roughly a factor of 75 over using Donaldson's algorithm alone.

Extended to the calculation of the metric itself, we predict a speed-up by a factor of 50 or so. If one wants to obtain high-precision Calabi--Yau metrics for scans over moduli or to check bundle stability, this factor of 50 is crucial, reducing the 50 or so hours of the $k=12$ calculation to a single hour. Checking 100 or even 1,000 different combinations of complex structure, K\"ahler and bundle moduli is then possible in days rather than years. If one combines this speed-up with a high-performance computing cluster, one could easily imagine not only exploring the moduli space of a single Calabi--Yau, but scanning over different choices of Calabi--Yau (such as the examples of \cite{Braun:2005ux,Braun:2005nv,Anderson:2011ns,Anderson:2012yf,Anderson:2013xka}).

Since this is the first time machine-learning techniques have been applied to the geometry of Calabi--Yau spaces, it is not surprising that there are many directions for future work.
As the next step, one could repeat everything we have done for more complicated examples, such as the torus-fibred Schoen Calabi--Yau  threefold with $\pi_1(X)=\mathbb{Z}_3\times\mathbb{Z}_3$ that leads to a heterotic standard model~\cite{Braun:2005ux}. Note that as the Calabi--Yau becomes more complicated, the sections (the polynomials) that enter the calculation of the $T$-operator also increase in complexity and take longer to evaluate. Thanks to this, pushing Donaldson's algorithm to higher $k$ becomes even more costly than in the case of the Fermat quintic. This suggests that one will see an even greater relative increase in speed using our approach.

We also note that Donaldson's algorithm and the approach we have suggested are agnostic about the dimension of the Calabi--Yau manifold. In particular, everything can be carried over to the study of four-folds appearing in F-theory compactifications. Given recent progress on finding MSSM-like models within F-theory~\cite{Lin:2014qga,Cvetic:2015txa,MayorgaPena:2017eda,Cvetic:2018ryq,Cvetic:2019gnh,Taylor:2019wnm}, it may soon become important to be able to access the Calabi--Yau metrics in such models numerically. Unfortunately, the extra two dimensions lead to a huge increase in the number of points one needs to use -- roughly, a calculation with $10^6$ points on a threefold would need $10^8$ points on a fourfold to achieve the same accuracy. Given this, a factor of 50 speed-up may be essential for the calculation of the metric to even be feasible.

Finally, the techniques we have used in this paper are relatively straightforward when compared with the vast array of machine-learning technology available today. It is certainly likely that much of what we have done could be improved upon, increasing both the accuracy and the speed of our approach, for example by treating the components of the metric as a single object. In a different direction, one might try to combine the ``energy functional'' method advocated by Headrick and Nassar\cite{Headrick:2009jz} and the neural network approach of Comsa et al.~\cite{Comsa:2019rcz}, using the $\sigma$ measure as a loss function to be minimised. Given the ability of machine-learning techniques and neural networks to handle high-dimensional optimisation problems, this may give algebraic metrics that are more accurate than Donaldson's balanced metrics in less time than a brute-force optimisation. We hope to make progress on this in the future.

\section*{Acknowledgements}
Anthony Ashmore and Burt Ovrut are supported in part by research grant DOE No.~DE-SC0007901. Yang-Hui He would like to thank the Science and Technology Facilities Council, UK, for grant ST/J00037X/1. We thank Volker Braun for sharing the C\texttt{++} code from previous work on numerical Calabi--Yau metrics. Anthony Ashmore would also like to thank Merton College, Oxford, for support at the start of this project.

\appendix


\section{Donaldson's algorithm in detail}\label{app:donaldson}

In this appendix, we discuss in more detail how to implement Donaldson's algorithm numerically. To be completely concrete, we will focus on the Calabi--Yau threefold $Q$ known as the Fermat quintic.

\subsection{The Fermat quintic}

Recall that $Q$ is given by the zero locus of a homogeneous quintic polynomial in $\IP^4$ whose projective coordinates we will take to be $[z_0: z_1: z_2: z_3: z_4]$
\begin{equation}\label{defQ}
Q := \{  \sum_{i=0}^4 z_i^5 = 0 \} \subset \IP^4.
\end{equation}
With this specific choice, let us spell out Donaldson's algorithm in more detail.

Fixing a positive integer $k$, $\{s_\alpha\}$ can be chosen to be monomials of degree $k$ on $Q$. On $\IP^n$, finding all monomials of degree $k$ is a standard problem in combinatorics and amounts to choosing $k$ elements from $n+k-1$. Since the ambient space is $\IP^4$, there are $\binom{5+k-1}{k}$ ways of doing so.
Explicitly, for the first few values of $k$, the monomial bases are
\begin{equation}
\begin{array}{c|c|c}
k & N_k & \{s_\alpha\} \\ \hline
1 & 5 & z_{i = 0, \ldots, 4} \\ \hline
2 & 15 & z_i z_j \ , 0 \leq i \leq j \leq 4 \\ \hline 
3 & 35 & z_i z_j z_k \ , 0 \leq i \leq j \leq k \leq 4 \\ \hline
4 & 70 & z_i z_j z_k z_\ell \ , 0 \leq i \leq j \leq k \leq \ell \leq 4
\end{array}
\end{equation}
On $Q$ we need to impose the defining quintic equation \eqref{defQ} when we encounter variables of powers greater than or equal to 5. For example, we can choose to replace $z_0^5 \to - \sum_{i=1}^4 z_i^5$.
This amounts to a reduction of the number of independent monomials of degree $k \geq 5$.
In general, we have that
\begin{equation}\label{Nk}
N_k = \left\{
\begin{array}{lcl}
\binom{5+k-1}{k},& 0 < k \leq 4 \\ 
 \binom{5+k-1}{k} -  \binom{k-1}{k-5} ,& k \geq 5 \\
\end{array}
\right.
\end{equation}

Since the metric is a local quantity, we need to focus on particular affine patches of $Q$. Suppose we are in the $z_0 = 1$ patch. We can eliminate one of the remaining four coordinates, say, $z_1 = (-1 - z_2^5 - \ldots -z_4^5)^{1/5}$,  so that the good local coordinates are $(z_2, z_3, z_4)$, which we then set to be $x^a$. The holomorphic volume form is then
\begin{equation}\label{VolForm}
\Omega = \int_Q \frac{\dd z_1 \wedge \dd z_2 \wedge \dd z_3 \wedge \dd z_4}{ 1 + z_1^5+ z_2^5+ z_3^5+ z_4^5} = 
\frac{\dd z_2 \wedge \dd z_3 \wedge \dd z_4}{5 z_1^4} ,
\end{equation}
where the Griffith residue theorem is applied in the last equality upon integrating out $z_1$.

In general, we
\begin{enumerate}
	\item
	Work in the affine patch defined by $z_I$ for some $I=0, \ldots, 4$. For numerical stability, $z_I$ should have the largest norm of the $z_i$.\footnote{In other words, if one works in homogeneous coordinates, we take $z_I$ to be the coordinate with the largest norm and then divide the other coordinates by $z_I$.}
	\item
	Eliminate one of the four remaining variables $z_{J \neq I}$ by solving for $z_J$ using the defining equation of $Q$. The remaining three variables then constitute the ``good coordinates'' of the patch on $Q$ -- we denote these by $x^a$ for $a=1,2,3$. For numerical stability (cf.~\S 3.4.2 of \cite{Douglas:2006rr}), we eliminate the variable for which $|\partial Q/\partial z_J|$ is the largest.
\end{enumerate}

In practice, random points on $Q$ (and any other quintic) can be chosen using a method given in~\cite{Douglas:2006rr}. The idea is that instead of picking random points on $Q$, one picks random lines in $\mathbb{P}^{4}$ and intersects them with $Q$. This gives five points that lie on the line giving you five random points in $Q$. A line $L$ in $\mathbb{P}^{4}$ is
\begin{equation}
L\simeq\mathbb{P}^{1}\subset\mathbb{P}^{4}.
\end{equation}
The intersection of the line with the quintic determines five points $L\cap Q=\{5\,\text{points}\}$ whose coordinates can then be found by numerically solving a quintic equation in one variable.
\begin{enumerate}
	\item Explicitly, pick two distinct points in $\mathbb{P}^{4}$
	\begin{equation}
	p=[p_{0}:\ldots:p_{4}],\qquad q=[q_{0}:\ldots:q_{4}].
	\end{equation}
	The line $L$ is defined as
	\begin{equation}
	\begin{split}
	L\colon\mathbb{C}\cup\{\infty\} & \to\mathbb{P}^{4},\\
	t & \mapsto[p_{0}+q_{0}\,t,\ldots,p_{4}+q_{4}\,t].
	\end{split}
	\end{equation}
	\item
	The five intersection points $L\cap Q$ are the solutions to
	\begin{equation}
	Q\circ L(t)=Q(p_{0}+q_{0}\,t,\ldots,p_{4}+q_{4}\,t)=0.
	\end{equation}
	\item
	To generate the initial points in $\mathbb{P}^{4}$, we use uniformly distributed points on the unit sphere $\text{S}^{9}\subset\mathbb{C}^{5}$. To get the points on S$^{9}$, start with the unit hypercube $[-1,1]^{10}\subset\mathbb{R}^{10}$, then take points which lie inside the unit disk D$^{10}$ and project radially to $\partial\text{D}^{10}=\text{S}^{9}$.
	\item
	Choosing what one means by ``random'' points picks out a particular integration measure $\dd A$. This auxiliary measure does \emph{not} give the correct distribution of points on $Q$ (which are distributed according to $\dd\vcy{}$) but it is simple to produce points with respect to it. What is the auxiliary measure $\dd A$ in this case? We have picked lines uniformly with respect to the $\text{SU}(5)$ action on $\mathbb{P}^{4}$. Following \S 3.4.2 of \cite{Douglas:2006rr}, the expected distribution of lines is then
	\begin{equation}
	\langle L\rangle\sim\omega_{\text{FS}}^{3},
	\end{equation}
	where $\omega_{\text{FS}}$ is the K\"ahler form on $\mathbb{P}^4$ defined by the $\text{SU}(5)$-invariant Fubini--Study K\"ahler potential
	\begin{equation}
	K_{\text{FS}}=\frac{1}{\pi}\ln\sum_{i}|z_{i}|^2.
	\end{equation}
	Using the embedding $i\colon Q \to\mathbb{P}^{4}$, the auxiliary measure is then simply the pull back of the volume form defined by the Fubini--Study metric
	\begin{equation}
	\dd A=\langle Q\cap L\rangle\sim i^{*}(\omega_{\text{FS}}^{3}).
	\end{equation}
	Note that, in general, the symmetry of the ambient space ($\text{SU}(5)$ in this case) is not enough to fully determine the auxiliary measure. For more complicated threefolds (not simply quintics), one needs a more general invariant theory.
\end{enumerate}

We can now sample points on $Q$. We choose $N_p$ random points  $\{p_M\}$ for  $M=1, \ldots, N_p$ using the above strategy. The number $N_p$ needs to be rather large and, in practice, one needs $N_p \gg N_k^2$ points \cite{Douglas:2006hz,Braun:2007sn,Anderson:2010ke}.\footnote{A rule of thumb $N_p = 10\, N_k^2 + 50{,}000$ was used in \cite{Braun:2008jp}.} Note that this means one needs to take approximately 200,000 points for $k=5$, 8,000,000 points for $k=10$, and 500,000,000 points for $k=20$.
%
%
%
%

As noted above, the auxiliary measure $\dd A$ does not give the correct distribution of points on $Q$. To ensure that the random points are chosen in an unbiased way when numerically integrating over $Q$, we \emph{weight} each point $p_M$ with
\begin{equation}\label{w}
w_M = \left. \frac{\dd\vcy{}}{\dd A} \right|_{p_M} ,
\end{equation}
where $|_{p_M}$ denotes the quantity evaluated at the point $p_M$. In \cite{Douglas:2006rr}, this weight is called the \emph{mass}. The weight can be evaluated for each point as we know $\dd\vcy{}=\Omega\wedge\bar\Omega$ (given by the residue formula) and $\dd A$ (fixed by the Fubini--Study form above). One can then numerically integrate quantities over the threefold using
\begin{equation}
\int_Q \dd\vcy{}\, f =\int_Q \frac{\dd\vcy{}}{\dd A}\dd A\, f =\frac{1}{N_p}\sum_{M=1}^{N_p} w_M \,f|_{p_M}.
\end{equation}
Note that we get a numerical estimate of the integrated Calabi--Yau volume by taking $f=1$:
\begin{equation}\label{eq:sum_weights}
\vcy{}=\int_Q \dd\vcy{}=\frac{1}{N_p}\sum_{M=1}^{N_p} w_M.
\end{equation}

Let us describe explicitly how one calculates these weights.
\begin{enumerate}
	
	\item
	Suppose we are in the patch where $z_I = 1$ and we eliminate the coordinate $z_J$ from \eqref{VolForm} using the defining quintic equation. We then have
	\begin{equation}
	\Omega \wedge \Ob = 5^{-2} |z_J|^{-8} \dd^2 x_1 \wedge \dd^2 x_2 \wedge \dd^2 x_3 ,
	\end{equation}
	where $x_a$ are the three remaining good coordinates ($z_i$ with $i\neq I\neq J$) and $\dd^2 x$ is understood to be $\dd x \wedge \dd\bar x$.

	\item
	The measure $\dd A$ can be defined following \S 3.4.2 of \cite{Douglas:2006rr} as
	\begin{equation}
	\dd A = ( i^*   \omega_{\IP^4}^{\text{FS}})^3 ,
	\end{equation}
	where $i^*$ is the pull-back induced from the embedding $i \colon Q \hookrightarrow \IP^4$ of the quintic into $\IP^4$ and $\omega_{\IP^4}^{\text{FS}}$ is the K\"ahler form associated with the standard Fubini--Study metric on $\IP^4$:
	\begin{equation}
	g_{i\jb}^{\text{FS}} = \partial_i \partial_{\jb} K^{\text{FS}} , 
	\qquad
	K^{\text{FS}} = \frac{1}{\pi} \ln \sum_{i = 0}^4 |z_i|^2 .
	\end{equation}
	The map $i^*$ is defined by the Jacobian $J^i{}_a=\partial z_i/\partial x_a$ as
	\begin{equation}
	(i^*\omega_{\IP^4}^{\text{FS}} )_{a\bar b}
	=
	J^i{}_a  (\omega_{\IP^4}^{\text{FS}})_{i\bar j}  \bar{J}^{\bar j}{}_{\bar b} .
	\end{equation}
	In the above expression, $(\omega_{\IP^4}^{\text{FS}})_{i\bar j}$ is a $5 \times 5$ matrix and the Jacobian $J^i{}_a$ is $3 \times 5$ matrix (since there are three good coordinates $x_a$ on $Q$).
	Most of the Jacobian will be the identity matrix, as we shall see shortly. Note that, as $\omega_{\IP^4}^{\text{FS}}$ is a two-form, $( i^*   \omega_{\IP^4}^{\text{FS}})^3$ is simply the determinant of $(i^*\omega_{\IP^4}^{\text{FS}} )_{a\bar b}$ multiplied by the top-form $\dd^2 x_1 \wedge \dd^2 x_2 \wedge \dd^2 x_3$ (up to a numerical factor).
\end{enumerate}
In summary, the weight associated to a random point, whose $I$-th coordinate is equal to 1 (i.e.~in the affine patch of $z_I$) with $z_J$ eliminated using the defining equation, is
\begin{equation}\label{w_M}
w_M = 5^{-2} |z_J|^{-8} (\det (i^*\omega_{\IP^4}^{\text{FS}} )_{a\bar b})^{-1} .
\end{equation}
Note that the overall coefficient in this expression is unimportant as it simply rescales $\vcy{Q}$.

\begin{enumerate}
	\item 
	We can now evaluate elements of the monomial basis $\{s_\alpha\}$ at each point $p_M$ and obtain the $T$-operator. Choosing some initial invertible hermitian matrix $h^{\alpha \bb}$, the $T$-operator is
	\begin{equation}
	T(h)_{\alpha \bb} = \frac{N_k}{\vcy{}}
	\sum_{M=1}^{N_p} \frac{s_\alpha(p_M) \overline{s_\beta(p_M)}}
	{h^{\gamma \bar{\delta}} s_\gamma(p_M) \overline{s_\delta(p_M)} } w_M , 
	\quad
	\alpha, \bb = 1,\ldots,N_k,
	\end{equation}
	where $\vcy{}$ is computed numerically from summing the weights, as in \eqref{eq:sum_weights}.
	
	\item
	Set the new $h^{\alpha \bb}$ to be $(T_{\alpha \bb})^{-1}$ and iterate.
	We point out that in practice \cite{Braun:2007sn}, one actually computes the transpose of the inverse, i.e.
	\begin{equation}
	h^{\alpha \bb}_{\text{new}} = \left[ (T_{\alpha \bb})^{-1} \right]^\text{T}.
	\end{equation}
	This is because the inverse of $T$ is numerically a matrix $T^{-1}$ such that $(T^{-1})^{\gb \alpha} T_{\alpha\bb}=\delta^{\gb}_{\bb}$. The algorithm is insensitive to the initial choice of $\hmat$ and in practice fewer than 10 iterations are needed to converge to the balanced metric. We use 10 iterations for the $T$-operator for all calculations in this paper. The actual computation for the metric $\gk$ is as follows.
	\begin{enumerate}
		\item 
		From $h^{\alpha \bb}$ we obtain the K\"ahler form in terms of the coordinates $(z_0,\ldots, z_4)$ 
		of the ``ambient'' affine patch (with, for example, $z_0=1$) on $\IP^4$:
		\begin{align}
		K(z, \zb) = 
		\frac{1}{k \pi} \ln \sum_{\alpha, \bar{\beta} = 1}^{N_k} h^{\alpha \bb} s_{\alpha}(z) \sb_{\bb}(\zb)\;
		\Rightarrow\;
		\tilde{g}_{i\bar j}(z, \zb) = \partial_i \partial_{\jb} K .
		\end{align}
		
		\item
		We pull-back this metric via the immersion of the Fermat quintic polynomial to find the metric $g_{a \bar b}$ on $Q$.
		Suppose, without loss of generality, that we are in the patch $z_0=1$ and the good coordinates on $Q$ are $x_a=(z_2, z_3, z_4)$ with $z_1$ to be eliminated via
		\begin{equation}
		z_1^5  = -1 - \sum_{i=2}^4 z_i^5 \quad
		\Rightarrow \quad
		\frac{\partial z_1}{\partial z_i} = - \frac{z_i^4}{z_1^4}\quad \ i = 2,3,4 .
		\end{equation}
		The other situations of different $I$ and $J$ are simply permutations of the following discussion. As the metric is a tensor, the pull-back is given simply by multiplying with Jacobian $J^i{}_a$. Explicitly the Jacobian is
		\begin{equation}
		\nn
		J^i{}_a = \frac{\partial z_i}{\partial x_a}=
		\frac{\partial z_{i=0,\ldots,4}}{\partial z_{k = 2,3,4}}
		=
		\left(
		\begin{array}{c|c|ccc}
		0&-z_2^4/z_1^4 & 1 & 0 & 0 \\
		0&-z_3^4/z_1^4 & 0 & 1 & 0 \\
		0&-z_4^4/z_1^4 & 0 & 0 & 1 \\
		\end{array}
		\right).
		\end{equation}
		\item Finally, using the metric $g_{a \bar b}$ on $Q$, the K\"ahler form is
		\begin{align}
		\omega = \frac{\ii}{2} \sum_{a, \bar b}^3 g_{a \bar b}(x, \bar x) \dd x^a \wedge \dd \bar{x}^{\bar b}.
		\end{align}
	\end{enumerate}

	\item
	To check the accuracy of the numerical metrics, one can calculate the error measures $\sigma$, $\Vert R \Vert$ and $\Vert EH \Vert$. We first take a sample $N_t < N_p$ of test points and calculate the volumes
	\begin{equation}\label{eq:vols}
	\vcy{Q} = \frac{1}{N_t} \sum_{M=1}^{N_t} w_M , \qquad
	\vk{Q} = \frac{1}{N_t} \sum_{M=1}^{N_t} 
	\frac{\omega^3(p_M)}{\Omega(p_M)  \wedge \overline{\Omega(p_M)}} w_M .
	\end{equation}
	The accuracy measures can then be obtained numerically as
	\begin{align}
	\sigma &=  \frac{1}{N_t \vcy{Q}} \sum_{M=1}^{N_t}
	\left| 
	1 - \frac{\omega(p_M)^3 / \vk{Q}}{\Omega(p_M)  \wedge \overline{\Omega(p_M)} / \vcy{Q} }
	\right| w_M ,\label{sigmak-1}\\
	\Vert R \Vert&=\frac{\vk{Q}^{1/3}}{N_t \vcy{Q}}\sum_{M=1}^{N_t} 
	\frac{\omega^3(p_M)}{\Omega(p_M)  \wedge \overline{\Omega(p_M)}} |R(p_M)|\,w_M,\label{R_numeric}\\
	\Vert EH \Vert&=\frac{1}{N_t \vk{Q}^{2/3}}\sum_{M=1}^{N_t} 
	\frac{\omega^3(p_M)}{\Omega(p_M)  \wedge \overline{\Omega(p_M)}} |R(p_M)|\,w_M,\label{EH_numeric}
	\end{align}
	where $|R(p_M)|$ is the absolute value of the Ricci scalar evaluated at $p_M$, calculated as in Appendix B.
	%
	
	%
\end{enumerate}

\section{Efficient numerical calculation of \texorpdfstring{$\sigma$}{sigma} and \texorpdfstring{$R$}{R}}\label{app:numerical}

Previous work on numerical Calabi--Yau metrics have used implementations in C or C\texttt{++}. Instead, we used Mathematica. Our choice was guided by the ease with which new examples can be implemented and the growing suite of machine learning tools available within Mathematica.

Traditionally, a compiled language such as C is much faster than a symbolic language such as that offered by Mathematica. What is likely less well known is that Mathematica is built on a set of numerical libraries that make numerical matrix calculations extremely efficient if used correctly. In this appendix, we discuss how we implemented Donaldson's algorithm using Mathematica, achieving speeds similar to or exceeding previous C implementations. This relies on a rewriting of the K\"ahler metric and the Ricci tensor so that manipulations are carried out on numerical rather than symbolic tensors as soon as possible. On a dual-core laptop computer with 8 GB of RAM, for the Fermat quintic at $k=8$, we can compute the $h^{\alpha\bar{\beta}}$ matrix for 2,000,000 points with 10 iterations in approximately 40 minutes. We can compute the $\sigma$ and $\Vert R \Vert$ measures for 500,000 test points in approximately 16 and 26 minutes respectively. On a 36-core workstation, we can compute the $h^{\alpha\bar{\beta}}$ matrix for 2,000,000 points and 500,000 test points at $k=8$ in approximately 270s, while the $\sigma$ and $\Vert R \Vert$ measures take 160s and 300s respectively. Note that for this number of test points, calculations can use over 100GB of RAM and so must be batched. Note also that these calculations do not use the symmetry of the Fermat quintic to reduce the number of independent components of the $T$-operator (as was done in \cite{Douglas:2006rr} and \cite{Headrick:2009jz}) or the adaptive mesh introduced in \cite{Anderson:2011ed}.

For what follows we work with the original basis $\{s_\alpha\}$ of monomials and include $h^{\alpha\bar\beta}$ explicitly. In practice, since one can diagonalise a balanced metric, it is more efficient to move to an orthonormal basis of sections in which $h^{\alpha\bar\beta} = \delta^{\alpha\bar\beta}$. Quantities such as $h^{\alpha\bar\beta}s_\alpha \bar{s}_{\bar\beta}$ can then be evaluated by a single vector dot product which scales as $\mathcal{O}(N_k)$ rather than a vector-matrix-vector product which scales as $\mathcal{O}(N_k^3)$.

\subsection{Metric and \texorpdfstring{$\sigma$}{sigma} measure}\label{sec:metric_sigma}

Let $z^{i}$, $i=0,\ldots,4$, be homogeneous coordinates on the ambient space $\mathbb{P}^{4}$ and let $x^{a}$, $a=1,2,3$ be good coordinates on the Calabi--Yau. Given a K\"ahler potential on $\mathbb{P}^{4}$, the corresponding K\"ahler metric is
\begin{equation}
\tilde g_{i\bar{j}}=\partial_{i}\partial_{\bar{j}}K.
\end{equation}
Using the embedding $x^{a}=x^{a}(z)$, one can pull back this tensor to get the metric\footnote{For ease of notation, in this appendix we denote the components of the metric by $g_{a\bar b}$ and the metric as a matrix by $g$. We will denote the determinant of the metric by $\det g$.} $g$ on $X$
\begin{equation}
g_{a\bar{b}}  =J^{i}{}_{a}J^{\bar{j}}{}_{\bar{b}}\partial_{i}\partial_{\bar{j}}K,\qquad
\tilde{g}  =J^{\text{T}}\tilde g\bar{J},
\end{equation}
where the Jacobian $J$ is a function of $z^{i}$ and not $\bar{z}^{\bar{i}}$, and so it can be moved through the derivatives -- this is just the statement that $\tilde g_{i\bar{j}}$ transforms as an honest $(1,1)$ tensor. The ansatz for the K\"ahler potential on the ambient space is
\begin{equation}
K(z,\bar{z})=\frac{1}{k\pi}\ln\sum_{\alpha,\bar{\beta}=1}^{N_{k}}h^{\alpha\bar{\beta}}s_{\alpha}(z)\bar{s}_{\bar{\beta}}(\bar{z}).
\end{equation}
Taking the mixed second derivatives of this, one can rewrite the expression for the metric on the ambient space as
\begin{equation}
\tilde{g}_{i\bar{j}}=\frac{1}{k\pi}\left(K^{(0)}K_{i\bar{j}}^{(2)}-(K^{(0)})^{2}K_{i}^{(1)}\overline{K_{j}^{(1)}}\right),\label{eq:metric_expand}
\end{equation}
where
\begin{align}
(K^{(0)})^{-1} & =\ee^{k\pi K}=\sum_{\alpha,\bar{\beta}=1}^{N_{k}}h^{\alpha\bar{\beta}}s_{\alpha}\bar{s}_{\bar{\beta}},\\
K_{i}^{(1)} & =\sum_{\alpha,\bar{\beta}=1}^{N_{k}}h^{\alpha\bar{\beta}}\partial_{i}s_{\alpha}\bar{s}_{\bar{\beta}},\\
K_{i\bar{j}}^{(2)} & =\sum_{\alpha,\bar{\beta}=1}^{N_{k}}h^{\alpha\bar{\beta}}\partial_{i}s_{\alpha}\partial_{\bar{j}}\bar{s}_{\bar{\beta}}.
\end{align}
Note also that, as $h^{\alpha\bar{\beta}}$ is hermitian, one has
\begin{equation}
\overline{K_{i}^{(1)}}=h^{\alpha\bar{\beta}}s_{\alpha}\partial_{\bar{i}}\bar{s}_{\bar{\beta}}=K_{\bar{i}}^{(1)},\qquad\overline{K_{i\bar{j}}^{(2)}}=h^{\beta\bar{\alpha}}\partial_{j}s_{\beta}\partial_{\bar{i}}\bar{s}_{\bar{\alpha}}=K_{j\bar{i}}^{(2)}.
\end{equation}

In order to calculate the $\sigma$ measure, one needs the value of $\omega^{3}$ for each point $p$ on $X$. Up to a constant, $\omega^{3}$ can be obtained by taking the determinant of $g_{a\bar{b}}$, which in turn is fixed by the Jacobian $J$ and the ambient metric $\tilde g_{i\bar j}$. There are two obvious ways one might calculate the numerical values of $g_{a\bar{b}}$:
\begin{enumerate}
	\item $K$ is a function of $z^{i}$ and $\bar{z}^{\bar{i}}$. Compute the second derivatives of $K$ exactly using Mathematica to get an analytic expression for $\tilde g_{i\bar{j}}$, multiply by the relevant Jacobian factor $J$ (which depends on the choice of good coordinates on $X$) and then evaluate the resulting expression for each point $p$ on $X$.
	\item $\tilde g_{i\bar{j}}$ can be written in terms of $s_{\alpha}$, $\partial_{i}s_{\alpha}$ and their complex conjugates. Compute $\partial_{i}s_{\alpha}$ analytically using Mathematica and then evaluate, $s_{\alpha}$, $\partial_{i}s_{\alpha}$ and $J^i{}_a$ for each point $p$ on $X$. Reconstruct the value of $\tilde g_{i\bar{j}}$ at each point using (\ref{eq:metric_expand}) and then multiply by the relevant Jacobian factors to obtain the values of $g_{a \bar b}$.
\end{enumerate}
As might be expected, the first of these is extremely slow even for small values of $k$. The ambient metric $\tilde g_{i\bar{j}}$ is a rather complicated function of $z^{i}$ and $\bar{z}^{\bar{i}}$ constructed from symbolic outer products of the sections and their derivatives, and evaluating this function for each point $p$ on $X$ is time consuming. The second method is much quicker. Both $s_{\alpha}$ and $\partial_{i}s_{\alpha}$ are relatively simple functions of the coordinates which can be evaluated at each point quickly. Crucially, these are then numerical tensors, and so Mathematica can use efficient numerical linear algebra libraries to carry out the matrix multiplications and tensor products.

\subsection{Ricci scalar}\label{sec:R_calc}

Two more measures of convergence to the Ricci-flat metric are the $\Vert R \Vert$ and $\Vert EH \Vert$ measures (although, in this paper, we only calculate the first of these). For these, one needs to calculate the Ricci scalar at each point $p$ of $X$. As we already have the metric $g_{a\bar{b}}$ on $X$ from computing the $\sigma$ measure, we now need the Ricci tensor. The Ricci tensor is given by
\begin{equation}
R_{a\bar{b}}=\partial_{a}\partial_{\bar{b}}\ln\det g.
\end{equation}
Peeling off the Jacobians, we find
\begin{equation}
R_{a\bar{b}}=J^{i}{}_{a}\bar{J}^{\bar{j}}{}_{\bar{b}}\partial_{i}\partial_{\bar{j}}\ln\det g.
\end{equation}
Using the matrix identities
\begin{equation}
\partial(\ln\det\boldsymbol{X})=\tr(\boldsymbol{X}^{-1}\partial\boldsymbol{X}),\qquad\partial(\boldsymbol{X}^{-1})=-\boldsymbol{X}^{-1}\partial\boldsymbol{X}\boldsymbol{X}^{-1},
\end{equation}
we have
\begin{align}
\partial_{\bar{j}}\ln\det g & =\tr( g^{-1}\partial_{\bar{j}} g),\\
\partial_{i}\partial_{\bar{j}}\ln\det g & =\tr(\partial_{i}g^{-1}\partial_{\bar{j}}g+g^{-1}\partial_{i}\partial_{\bar{j}}g)\nonumber\\
& =\tr(-g^{-1}\partial_{i}g g^{-1}\partial_{\bar{j}}g+g^{-1}\partial_{i}\partial_{\bar{j}}g).
\end{align}
Given that $g=J^{\text{T}}\tilde g\bar{J}$ and that $J$ is a function of the $z^{i}$ only, we also have
\begin{align}
\partial_{i}g & =\partial_{i}J^{\text{T}}\tilde g\bar{J}+J^{\text{T}}\partial_{i}\tilde g\bar{J},\\
\partial_{i}\partial_{\bar{j}}g & =\partial_{i}J^{\text{T}}\partial_{\bar{j}}\tilde g\bar{J}+J^{\text{T}}\partial_{i}\partial_{\bar{j}}\tilde g\bar{J}+\partial_{i}J^{\text{T}}\tilde g\partial_{\bar{j}}\bar{J}+J^{\text{T}}\partial_{i}\tilde g\partial_{\bar{j}}\bar{J}.
\end{align}
We already have $\tilde{g}$, $g$ and $J$ (plus the conjugates and transposes) from the previous calculation of the $\sigma$ measure. We then need to compute $\partial_{i}J$, $\partial_{i}\tilde g$ and $\partial_{i}\partial_{\bar{j}} \tilde g$. The first of these, $\partial_{i}J$, is simple for Mathematica to compute analytically. The others are more complicated and so we would like to reduce them to derivatives of the sections $s_{\alpha}$. Note that we do not need to compute $\partial_{\bar{i}}\tilde g$ independently as it is determined from $\partial_{i}g$ as $\partial_{\bar{i}} \tilde g_{j\bar{k}}=(\partial_{i}\tilde g_{k\bar{j}})^{*}$ for a K\"ahler metric.

Using the ansatz for the K\"ahler potential, one can expand $\partial_{i} \tilde g$ to give
\begin{align}
k\pi\partial_{i}\tilde g_{k\bar{l}} & =\partial_{i}\bigl(K^{(0)}K_{k\bar{l}}^{(2)}-(K^{(0)})^{2}K_{k}^{(1)}\overline{K_{l}^{(1)}}\bigr)\\
& =-(K^{(0)})^{2}(K_{i}^{(1)}K_{k\bar{l}}^{(2)}+K_{k}^{(1)}K_{i\bar{l}}^{(2)}+\overline{K_{l}^{(1)}}K_{ik}^{(2)})+K^{(0)}K_{ik\bar{l}}^{(3)}+2(K^{(0)})^{3}K_{i}^{(1)}K_{k}^{(1)}\overline{K_{l}^{(1)}},\nonumber
\end{align}
where we have used
\begin{align}
\partial_{i}K^{(0)} & =-(K^{(0)})^{2}h^{\alpha\bar{\beta}}\partial_{i}s_{\alpha}\bar{s}_{\bar{\beta}} & \partial_{i}\overline{K_{l}^{(1)}} & =h^{\alpha\bar{\beta}}\partial_{i}s_{\alpha}\partial_{\bar{l}}\bar{s}_{\bar{\beta}}\nonumber\\
& =-(K^{(0)})^{2}K_{i}^{(1)}, &  & =K_{i\bar{l}}^{(2)},\\
\partial_{i}K_{k}^{(1)} & =h^{\alpha\bar{\beta}}\partial_{i}\partial_{k}s_{\alpha}\bar{s}_{\bar{\beta}} & \partial_{i}K_{k\bar{l}}^{(2)} & =h^{\alpha\bar{\beta}}\partial_{i}\partial_{k}s_{\alpha}\partial_{\bar{l}}\bar{s}_{\bar{\beta}},\nonumber\\
& \equiv K_{ik}^{(2)}, &  & \equiv K_{ik\bar{l}}^{(3)}.
\end{align}
Note also that $\partial_{i}K_{k}^{(1)}=\partial_{j}K_{i}^{(1)}$ and $\partial_{i}K_{k\bar{l}}^{(2)}=\partial_{k}K_{i\bar{l}}^{(2)}$.

The second derivatives $\partial_{i}\partial_{\bar{j}} \tilde g$ can also be written in terms of $s_{\alpha}$ and their derivatives as
\begin{align}
k\pi\partial_{i}\partial_{\bar{j}}\tilde g_{k\bar{l}} & =\partial_{i}\Bigl(\partial_{\bar{j}}K^{(0)}K_{k\bar{l}}^{(2)}+K^{(0)}\partial_{\bar{j}}K_{k\bar{l}}^{(2)}-2\partial_{\bar{j}}K^{(0)}K^{(0)}K_{k}^{(1)}\overline{K_{l}^{(1)}}\nonumber\\
& \eqspace\phantom{\partial_{i}\Bigl(}-(K^{(0)})^{2}\partial_{\bar{j}}K_{k}^{(1)}\overline{K_{l}^{(1)}}-(K^{(0)})^{2}K_{k}^{(1)}\partial_{\bar{j}}\overline{K_{l}^{(1)}}\Bigr)\nonumber\\
& =K^{(0)}K_{ik\bar{j}\bar{l}}^{(4)}-(K^{(0)})^{2}(K_{i\bar{j}}^{(2)}K_{k\bar{l}}^{(2)}+K_{ik}^{(2)}\overline{K_{jl}^{(2)}}+K_{k\bar{j}}^{(2)}K_{i\bar{l}}^{(2)})\nonumber\\
& \eqspace-(K^{(0)})^{2}(\overline{K_{j}^{(1)}}K_{ik\bar{l}}^{(3)}+\overline{K_{l}^{(1)}}K_{ik\bar{j}}^{(3)}+K_{i}^{(1)}\overline{K_{jl\bar{k}}^{(3)}}+K_{k}^{(1)}\overline{K_{jl\bar{i}}^{(3)}})\nonumber\\
& \eqspace+2(K^{(0)})^{3}(K_{i}^{(1)}\overline{K_{j}^{(1)}}K_{k\bar{l}}^{(2)}+K_{i\bar{j}}^{(2)}K_{k}^{(1)}\overline{K_{l}^{(1)}}+\overline{K_{j}^{(1)}}K_{k}^{(1)}K_{i\bar{l}}^{(2)}\nonumber\\
& \eqspace\phantom{+2(K^{(0)})^{3}(}+K_{i}^{(1)}K_{k\bar{j}}^{(2)}\overline{K_{l}^{(1)}}+K_{i}^{(1)}K_{k}^{(1)}\overline{K_{jl}^{(2)}}+\overline{K_{j}^{(1)}}K_{ik}^{(2)}\overline{K_{l}^{(1)}})\nonumber\\
& \eqspace-6(K^{(0)})^{4}K_{i}^{(1)}\overline{K_{j}^{(1)}}K_{k}^{(1)}\overline{K_{l}^{(1)}},
\end{align}
where we have used
\begin{align}
\partial_{i}\partial_{\bar{j}}K^{(0)} & =-2\partial_{i}K^{(0)}K^{(0)}\overline{K_{j}^{(1)}}-(K^{(0)})^{2}\partial_{i}\overline{K_{j}^{(1)}} & \partial_{i}\partial_{\bar{j}}K_{k}^{(1)} & =\partial_{i}K_{k\bar{j}}^{(2)}\nonumber\\
& =2(K^{(0)})^{3}K_{i}^{(1)}\overline{K_{j}^{(1)}}-(K^{(0)})^{2}K_{i\bar{j}}^{(2)}, &  & =K_{ik\bar{j}}^{(3)},\\
\partial_{i}\partial_{\bar{j}}K_{k\bar{l}}^{(2)} & =h^{\alpha\bar{\beta}}\partial_{i}\partial_{k}s_{\alpha}\partial_{\bar{j}}\partial_{\bar{l}}\bar{s}_{\bar{\beta}} & \partial_{i}\partial_{\bar{j}}\overline{K_{l}^{(1)}} & =h^{\alpha\bar{\beta}}\partial_{i}s_{\alpha}\partial_{\bar{j}}\partial_{\bar{l}}\bar{s}_{\bar{\beta}}\nonumber\\
& =K_{ik\bar{j}\bar{l}}^{(4)}, &  & =\overline{K_{jl\bar{i}}^{(3)}}.
\end{align}

As with the $\sigma$ measure, computing the Ricci tensor symbolically and evaluating it for each point $p$ is extremely slow. Instead, we calculate the derivatives of $s_{\alpha}$, evaluate them for each point $p$ on $X$ and then use the efficient numerical linear algebra routines available in Mathematica to reconstruct the Ricci tensor from the various $K^{(p)}$ tensors we have defined. Given the Ricci tensor, tracing with the metric $g$ gives the desired Ricci scalar.


\section{More on machine learning}\label{sec:gbt}

The method we have used to perform the machine learning in this paper is \emph{gradient-boosted decision trees}.
This was chosen after a comparison of performance with the various standard techniques such as support vector machines, nearest neighbors, neural networks and even simple linear regression. It was found that decision trees were by far the best for predicting the determinant. In this appendix, we will give a rapid introduction to this machine-learning method for non-experts; for the interested reader, further details can be found in \cite{mohri2018foundations,hartshorn2016machine}. We also provide more checks of our supervised-learning routine, including the training curves.

\subsection{Decision trees}
Suppose we have a target variable $y$ and a set of input variables $x_i$. Typically, $y$ is \emph{discrete}.
However, we can treat a continuous variable as discrete by splitting it into appropriate intervals and taking the average of the interval to be its discretized value. Our primary target variable in this paper is the determinant of the metric, $\detg k$ (note that this is clearly a continuous real variable and we will accordingly discretise it). The input variables are: (a) the complex affine coordinates of the point on the quintic at which $\detg k$ is to be computed; (b) the values of $\detg k$ at each point for some low values of $k$, say, $\detg 4$; (c) the values of $\detg k$ at each point for the desired higher value of $k$, say, $\detg{12}$.

A tree is then built from the input variables as a sequence of if/then/else statements.
Starting with $x_1$, we create a node and then travel down a ``branch'' depending on the value of $x_1$: if $x_1 \in [a_1, b_1]$ then we proceed to the first branch, 
if $x_1 \in [b_1, c_1]$, we proceed to a different branch, etc. At the end of each branch we create new node, say for $x_2$, from which new branches are created. As before, we then travel down one of these new branches depending on the value of $x_2$. Repeating this for all of the input variables gives a set of nodes partially connected by branches, giving a tree structure. The outermost nodes (the leaves) correspond to different predictions for the value of the output variable $y$. Once the tree structure is established, we can present it with a new input $x_i$ and then follow the tree structure to find the predicted output (the leaf) that it leads to. 

As in regression, the optimal parameters $(a_j, b_j, c_j \ldots)$ are determined by optimizing some goodness-of-fit score, for example the sum of squares of the error between the actual and predicted target variable. To prevent over-fitting, one often sets a maximum depth for the tree structure or a maximum number of leaves.

In a way, a decision tree can be understood as a highly non-analytic analogue of regression: whereas regression ultimately fits the target into some differentiable function $y = f(x_i)$ by minimizing sum squared errors, a decision tree writes down $y$ from $x_i$ via a sequence of discrete choices. Thus, when the output data $y$ is highly fluctuating with respect to $x_i$ (as in the case of $\detg k$), regression is not so useful since an analytic function $f$ is difficult to find; slotting into a sequence of decisions is much more appropriate.

\subsection{Ensembles and gradient boosting}
In order to improve the performance of predictions via decision trees, one can set up an ensemble (or forest) of trees, each with different decision criteria. The overall score of the prediction can be taken to be the sum over all trees.

Now, optimizing all parameters in all trees at once could become computationally intractable. Instead, one can take an additive strategy so that the predicted target $y$ is obtained by adding one tree at a time:
$y^{(j)} = y^{(j-1)} + g_j(x_i)$ where $g_j$ heuristically represents the function which captures the information about the tree at stage $j$.
The ensemble of trees is thus ``boosted'' iteratively and the tree to add at each stage is simply the one which optimizes the overall fitness score.

\subsection{Training curves and learning higher \texorpdfstring{$k$}{k} from \texorpdfstring{$k=1$}{k=1}}


Let us examine the training curves for some of the supervised-learning models we have discussed in the main text. We fix 10,000 samples of data $\cD_{1,k}$ on the quintic. As discussed in the main text, our strategy is to see how well the ML predicts $\detg k$ for $k=2,3,4,\ldots$ by using the predicted values to compute the $\hat\sigma$ measure. We can then compare this with the values of $\sigma$ computed using the actual values of $\detg k$ from Donaldson's balanced metric.

Consider $k=2$ first. To examine the learning curve, we split the data into a training set $\cT$ and a validation set $\cV$, and then vary the size of $\cT$. We train the ML on $\cT$ and then \emph{predict} the values of $\detg 2$ for both the training and validation sets. We then compute the $\sigma$-measure for the entire 10,000 points to see how well the predicted values reproduce the determinant of the balanced metric at $k=2$.\footnote{Note that in computing $\sigma$ we mix the training set and the validation set -- this is different from standard cross-validation where the goodness-of-fit is computed only for the validation set. We do this because $\sigma$ is a global quantity and we are integrating over as many points as possible.}

\begin{figure}[!h!t!b]
	\centerline{
		\includegraphics[trim=0mm 0mm 0mm 0mm, clip, width=10cm]{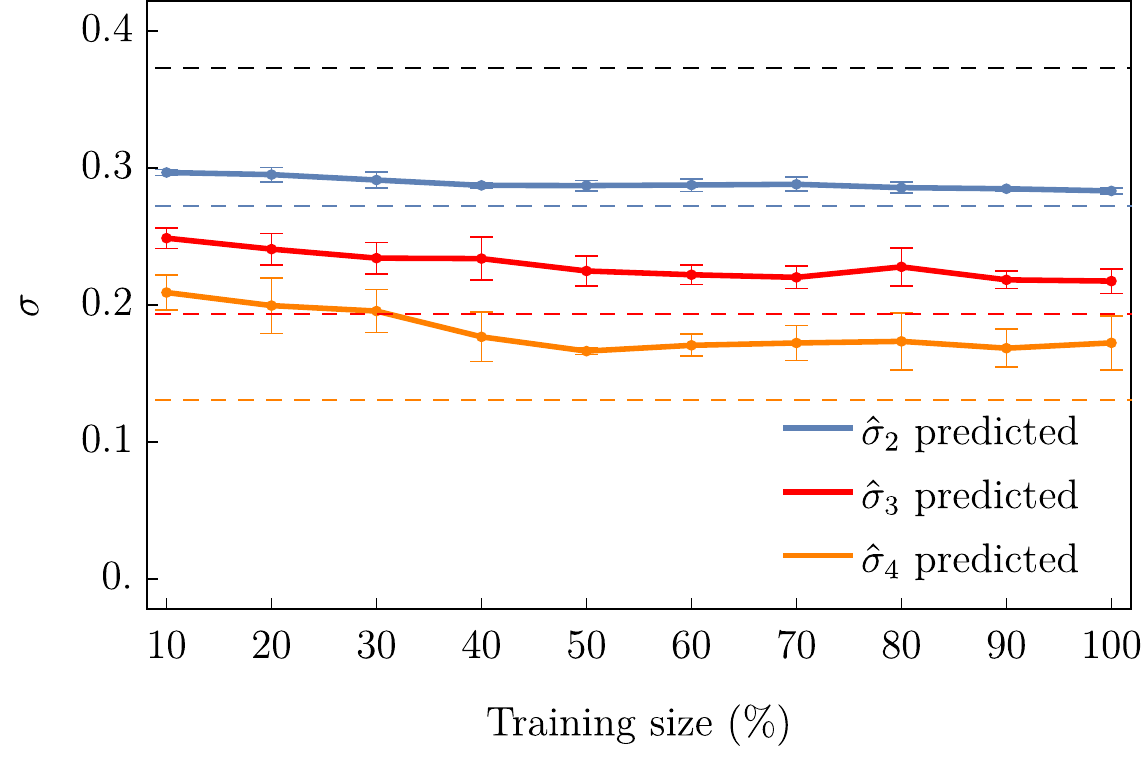}
	}
	\caption{Learning curves for a dataset of 10,000 random points for $\detg k$, learning from $k=1$ and predicting the $\hat\sigma_k$ results for $k=2,3,4$. As a comparison, the dashed lines denote the values of $\sigma_k$ computed using Donaldson's balanced metrics (black for $k=1$, blue for $k=2$, red for $k=3$ and orange for $k=4$.).
		\label{f:TCdetgk=1tok=n}}
\end{figure}

We plot the predicted $\hat\sigma$ for $k=2$ against the size of the training set in blue in Figure \ref{f:TCdetgk=1tok=n} -- this is the learning curve of the ML. Note that we take ten different random samples of each percentage, thus there is an error bar attached to each point. We see that even at 10\%, the $\hat\sigma$ value is already quite close the true value for $k=2$, around 0.3, decreasing steadily as we increase training size. Encouragingly, the curve is rather flat -- adding more data to the training set does not significantly improve the predicted value of $\hat\sigma$. Conversely, this means that we need to train on only a small number of $\detg 2$ values to accurately predict the remaining ones.

We repeat this procedure for higher values of $k$, trying to predict the determinant for $k=3$ and $k=4$ from $k=1$ given only some of the $k=3$ and $k=4$ results. The corresponding training curves are shown in red and orange in Figure  \ref{f:TCdetgk=1tok=n}. Each training curve (for $k=1 \to 2$, $k = 1 \to 3$ and $k=1 \to 4$) decreases as the size of the training set increases, meaning that the $\sigma$ measure is getting smaller (and closer to the actual value computed from the balanced metric) as the ML is trained on a greater number of ``correct'' values, as one would expect. Note that the curves shift down when the target value of $k$ is higher -- at higher $k$ the ML is learning a better approximation to the honest Ricci-flat metric.

\begin{figure}[!h!t!b]
	\centerline{
		\hspace{-5em}\includegraphics[trim=0mm 0mm 0mm 0mm, clip, width=10cm]{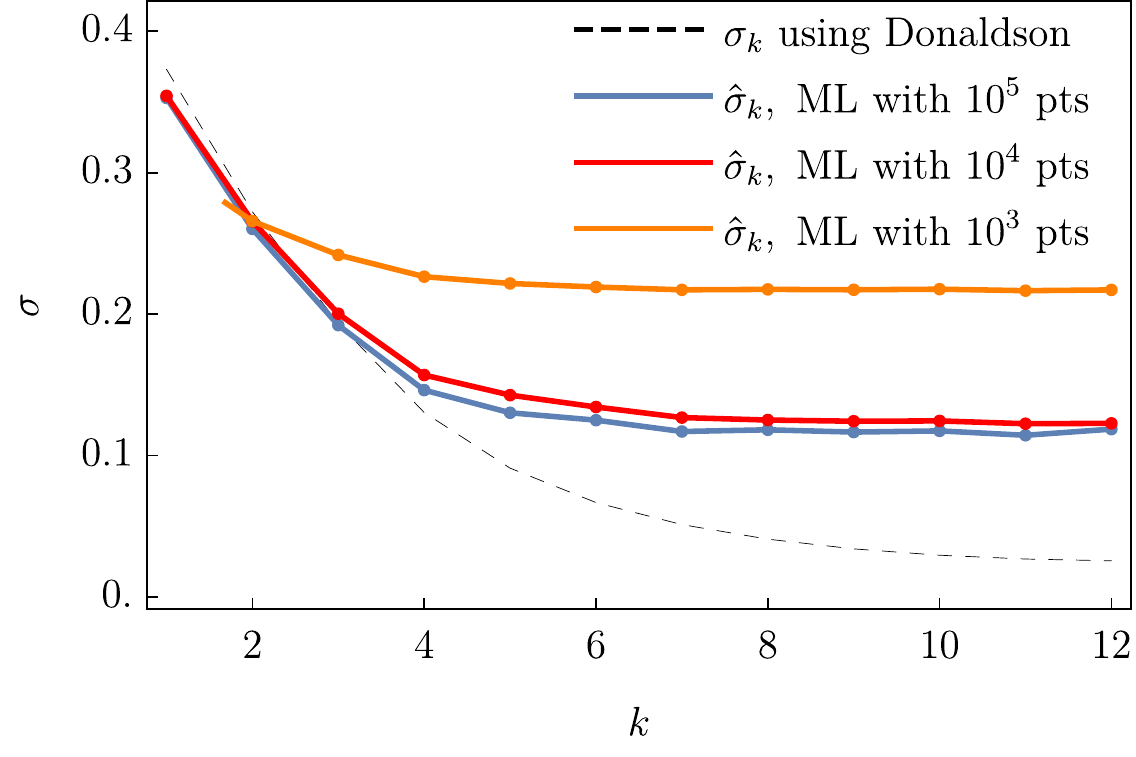}
	}
	\caption{Comparison of the $\sigma$ measure calculated using the known $\detg k$ values from the balanced metric and the values predicted using only the random points as input. The blue curve corresponds to $\sigma_k$ computed using Donaldson's balanced metrics. The remaining curves are $\hat\sigma_k$ for varying sizes of training data ($10^3$, $10^4$ and $10^5$ training samples). We see the accuracy (as measured by $\sigma$) improves with more training data up to a point.
		\label{fig:points_only}}
\end{figure}

Out of interest, we also trained MLs to predict the values of $\detg k$ using only the random points $\{p\}$ as inputs, that is, without any determinant data. The result of this is shown in Figure \ref{fig:points_only}. We varied the size of the training data from $10^3$ to $10^5$ and computed the $\sigma$ error measure using 10,000 validation points. We see that the ML is worse when using only the points as an input, particularly when trying to predict $\detg k$ for larger values of $k$. As $k$ increases, the functions that appear in the metric greatly increase in complexity, so it is not surprising that this simple ML has trouble finding the relation between the points and $\detg k$ for higher values of $k$. In particular, it appears that there is not sufficient information in the points data alone for our relatively simple ML to learn $\detg k$ past $k\approx 4$.

\subsection{The necessity of machine learning}

It may have occurred to the reader that since we have numerically generated the random points and associated determinant of the metric, of the form
\begin{equation}
\{p\}\to\{g|_p\}
\end{equation}
for hundreds of thousands of points $p$ on $Q$, why use ML at all? Why not simply perform an appropriately clever (non-linear) \emph{regression}? After all, if the points were simply on $\IP^4$, we might have guessed a function like $\log\sum_{i=0}^4 |z_i|^2$.

However, from the experience of computing cohomology groups \cite{He:2017aed,He:2018jtw,Larfors:2019sie,Constantin:2018hvl,Brodie:2019dfx,Donagi:2004qk}, whilst there might be a relatively simple formula on the ambient space, the restriction even to a hypersurface produces many subtleties in dividing regions where the ranks of the cohomology groups jump. Similarly, we expect the metric and K\"ahler potential to be complicated functions of the variables. Even though we can write, order-by-order in the degree $k$, the K\"ahler potential as a high-degree polynomial in the coordinates $z_i$ and we know from Donaldson's theorem that it converges to {\it some} function as $k \to \infty$, the precise functional form is unknown analytically.

\begin{figure}[!h!t!b]
	\centering
	\begin{subfigure}{.5\textwidth}
		\centering
		\includegraphics[width=7cm]{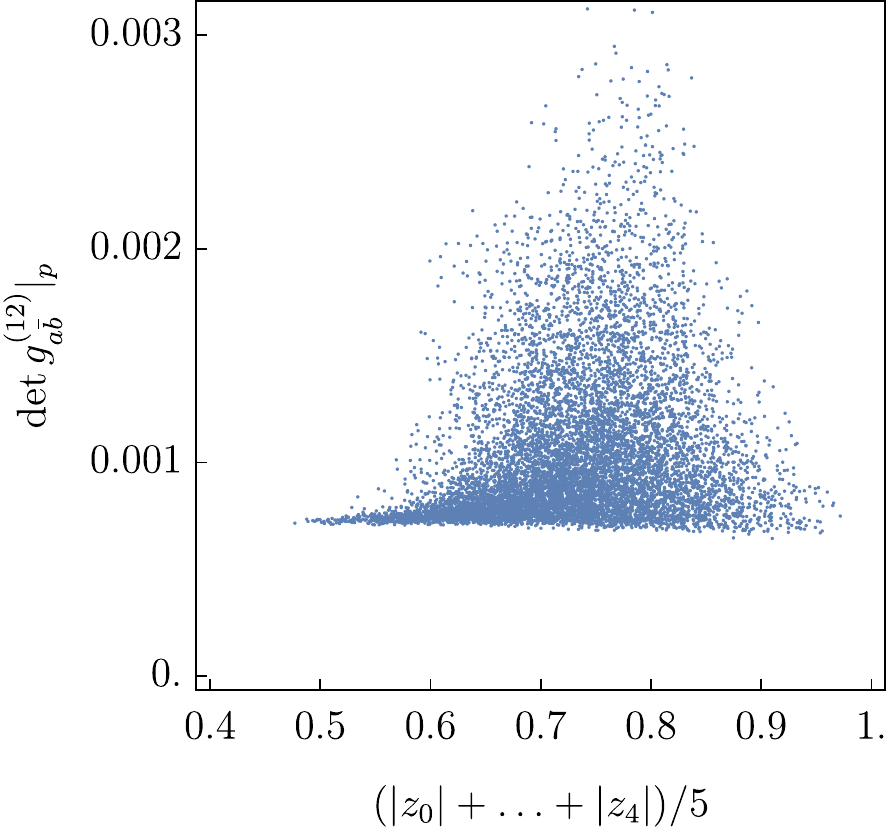}
		\caption{}

	\end{subfigure}%
	\begin{subfigure}{.5\textwidth}
		\centering
		\includegraphics[width=7cm]{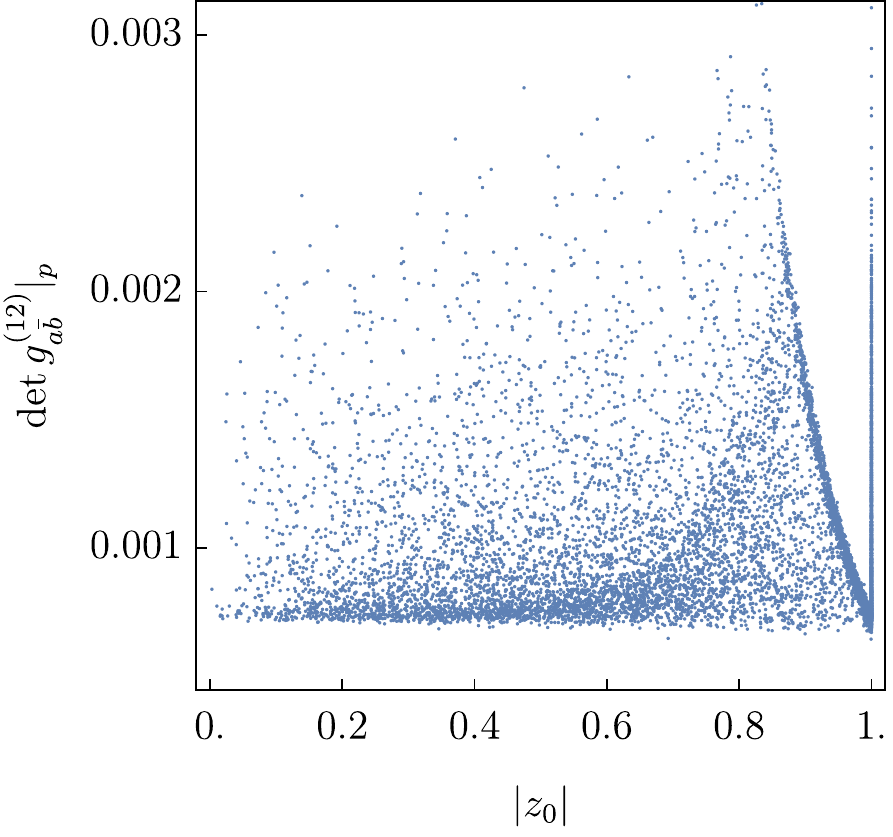}
		\caption{}

	\end{subfigure}
	\caption{Scatter plots of $\det g_{a\bar b}^{(12)}$ against (a) the mean modulus of the coordinates $(|z_0|+\ldots+|z_4|)/5$ and (b) the modulus of $z_0$.}
		\label{f:plot-g}
	
\end{figure}


To give an idea of how complicated this function is, let us plot $\det g_{a \bar b}^{(k)}$ at a high value (say $k=12$) against: (a) the mean of the modulus of all the variables, that is, 
$(|z_0|+\ldots+|z_4|)/5$; and (b) the modulus of a chosen variable, say $z_0$.
These are shown in parts (a) and (b) of Figure \ref{f:plot-g} respectively.
It is evident that $\det g_{a \bar b}^{(12)}$ is no simple function of the coordinates! Regression is useful only if one has some intuition about the functional form of what one is trying to reproduce.
In the absence of this intuition, one needs to turn to machine learning.

\bibliographystyle{utphys}
\bibliography{citations}

\end{document}